\documentclass[fleqn,usenatbib]{mnras}

\usepackage{newtxtext,newtxmath}
\usepackage[T1]{fontenc}
\usepackage{multicol}
\usepackage{graphicx}	
\usepackage{amsmath}	
\usepackage{hyperref}
\usepackage[dvipsnames]{xcolor}
\usepackage[utf8x]{inputenc}
\DeclareRobustCommand{\VAN}[3]{#2}
\let\VANthebibliography\thebibliography
\def\thebibliography{\DeclareRobustCommand{\VAN}[3]{##3}\VANthebibliography}

\newcommand{\bea}{\begin{eqnarray}}
\newcommand{\be}{\begin{equation}}
\newcommand{\ben}{\begin{enumerate}}
\newcommand{\bi}{\begin{itemize}}
\newcommand{\eea}{\end{eqnarray}}
\newcommand{\ee}{\end{equation}}
\newcommand{\ei}{\end{itemize}}
\newcommand{\een}{\end{enumerate}}

\title[Multi-probe forecast for CSST]{Cosmological Constraint Precision of the Photometric and Spectroscopic Multi-probe Surveys of China Space Station Telescope (CSST)}

\author[Miao H. et al.]{
Haitao Miao$^{1,2}$, 
Yan Gong$^{1,2}$\thanks{E-mail: gongyan@bao.ac.cn}, 
Xuelei Chen$^{3,4,5,6}$, 
Zhiqi Huang$^{7,8}$, 
Xiao-Dong Li$^{7,8}$, 
Hu Zhan$^{1,9}$
\\
$^{1}$key laboratory of space astronomy and technology, National Astronomical Observatories, \\Chinese Academy of Sciences, Beijing 100101, People's Republic of China\\
$^{2}$Science Center for China Space Station Telescope, National Astronomical Observatories, \\Chinese Academy of Sciences, Beijing 100101, People's Republic of China\\
$^{3}$Key Laboratory for Computational Astrophysics, National Astronomical Observatories, \\Chinese Academy of Sciences, 20A Datun Road, Beijing 100101, China \\
$^{4}$University of Chinese Academy of Sciences, Beijing 100049, People's Republic of China\\
$^{5}$Centre for High Energy Physics, Peking University, Beijing 100871, People's Republic of China\\
$^{6}$Department of Physics, College of Sciences, Northeastern University, Shenyang 110819, China\\
$^{7}$School of Physics and Astronomy, Sun Yat-sen University, 2 Daxue Road, Tangjia, Zhuhai, 519082, China\\
$^{8}$CSST Science Center for the Guangdong-Hongkong-Macau Greater Bay Area, SYSU, China\\
$^{9}$Kavli Institute for Astronomy and Astrophysics, Peking University, Beijing 100871, China
}

\date{Accepted XXX. Received YYY; in original form ZZZ}

\pubyear{2022}

\begin{document}
\label{firstpage}
\pagerange{\pageref{firstpage}--\pageref{lastpage}}
\maketitle

\begin{abstract}
 As one of Stage IV space-based telescopes, China Space Station Telescope (CSST) can perform photometric and spectroscopic surveys simultaneously to efficiently explore the Universe in extreme precision. In this work, we investigate several powerful CSST cosmological probes, including cosmic shear, galaxy-galaxy lensing, photometric and spectroscopic galaxy clustering, and number counts of galaxy clusters, and study the capability of these probes by forecasting the results of joint constraints on the cosmological parameters. By referring to real observational results, we generate mock data and estimate the measured errors based on CSST observational and instrumental designs. To study the systematical effects on the results, we also consider a number of systematics in CSST photometric and spectroscopic surveys, such as the intrinsic alignment, shear calibration uncertainties, photometric redshift uncertainties, galaxy bias, non-linear effects, instrumental effects, etc. The Fisher matrix method is used to derive the constraint results from individual or joint surveys on the cosmological and systematical parameters. We find that the joint constraints by including all these CSST cosmological probes can significantly improve the results from current observations by one order of magnitude at least, which gives $\Omega_m$ and $\sigma_8$  $<$1\% accuracy, and $w_0$ and $w_a$ $<$5\% and 20\% accuracies, respectively. This indicates that the CSST photometric and spectroscopic multi-probe surveys could provide powerful tools to explore the Universe and greatly improve the studies of relevant cosmological problems.
\end{abstract}


\section{Introduction}

Based on a variety of observations in the past two decades \citep{Percival10,Eisenstein2011AJ,Komatsu2011ApJS,Beutler11,Sievers13,Hinshaw13,Hou14,Planck13},
we have entered the era of precision cosmology, and established the standard cosmological model. 
Current observations indicate that the Universe is composed of about 70\% dark energy and 25\% dark matter. Dark energy provides a physical mechanism for cosmic acceleration, and it is characterized by its equation of state (EOS) $w$, which is defined by the ratio of pressure to energy density. When $w=-1$, dark energy becomes the cosmological constant, that is a key component of the standard Lambda cold dark matter ($\Lambda$CDM) model. The $\Lambda$CDM model is supported by most of current observations \citep[e.g.][]{Planck18,Alam21}. Although some tensions or challenges are still existing, such as missing satellites, cusp/core, and Hubble tension \citep[see e.g.][]{Bullock2017,Verde2019,Valentino2021}, the $\Lambda$CDM model has achieved tremendous success. However, the nature of dark energy and dark matter is still poorly understood. We still face great challenges of finding a fundamental physical explanation for dark energy and dark matter. Dynamical dark energy models have been studied extensively, where the EOS is usually time-dependent in these models. The most popular model is the Chevallier-Polarski-Linder parameterization \citep[CPL,][]{Chevallier01,Linder03}, which is a Taylor expansion around $a = 1$, $w = w_0 + w_a(1 - a)$. A number of cosmological observations can provide valuable information to explore the nature of dark energy and dark matter with high precision, such as the surveys related to the cosmic large scale structure (LSS). 

Galaxies formation is mainly caused by the collapse of matter under the effect of gravity. The spatial distribution of galaxies in the Universe contains a lot of information about the formation and evolution of cosmic structures. Galaxy catalogs with spectroscopic redshift are used to probe the three-dimensional structure of the Universe. The galaxy two-point spatial correlation function or power spectrum (i.e. Fourier transformation of the two-point correlation function) is often adopted in the data analysis. It can provide tight constraints on cosmological parameters by using only Baryon Acoustic Oscillation (BAO) wiggles \citep{Seo2007,Ribera2014} or full shape of galaxy power spectrum \citep{Wang2013MNRAS,Amico2020,Ivanov2020,Brieden2021JCAP}.

On the other hand, although photometric observations have a large redshift uncertainty, they can detect and measure a large number of galaxies than spectroscopic observations. We can extract accurate information from the galaxy two-point angular correlation function or angular power spectrum, which can improve the ability of constraining the cosmological parameters. The results from ongoing and near-future photometric surveys have been studied \citep[e.g.][]{Zhan2006b,Padmanabhan2007,Zhan2009,DES2021bao,Rosell2022,Monroy2022}. It has been shown that weak lensing is a precise cosmological probe by measuring shear information of galaxy images \citep{zhan06,Albrecht06,Kilbinger15,Hildebrandt17,Troxel2018,Amon2022PhRvD}. When the cross-correlation of the galaxy number density and galaxy shear is considered, more useful information and stringent constraint results can be obtained \citep{Zhang2010,Benjamin2010,McQuinn2013,Prat2018,Schaan20,Prat2021}. Hence, a joint analysis of  weak lensing, galaxy clustering and galaxy-galaxy lensing (the so-called 3$\times$2pt probes) have been suggested and applied to observational data \citep[e.g.][]{Bernstein09,Joachimi10,Uitert2018MNRAS,Krause17a,Krause17b,Eifler21a,Euclid20,Joachimi21,Krause21}. 
The results from Kilo-Degree Survey\footnote{http://www.astro-wise.org/projects/KIDS/} \citep[KiDS,][]{Hildebrandt17,Heymans21,Troster21}, Dark Energy Survey\footnote{www.darkenergysurvey.org/} \citep[DES,][]{Abbott18,Krause21,To21b,Abbott2022}, and Hyper Suprime-Cam\footnote{http://www.naoj.org/Projects/HSC/HSCProject.html} \citep[HSC,][]{Hikage21} have demonstrated the approach of increasing the overall constraining power by jointly analyzing different cosmological probes. 

Besides, number counts of galaxy clusters is also a powerful cosmological probe \citep[e.g.,][]{Yoo12,To21a,Eifler21b}. 
Galaxy clusters are formed at high peak of the primary matter density field. The abundance of galaxy clusters is sensitive to the growth history of structures and the expansion history of the Universe. So the number counts of galaxy clusters have a strong dependence on several cosmological parameters, especially the amplitude parameter of the matter power spectrum $\sigma_8$ and the matter energy density parameter of the Universe $\Omega_{\rm m}$. The cluster number density evolving with redshift can break the degeneracy between these two parameters, and thus can be used to further constrain the cold dark matter and dark energy density parameters \citep{Haiman2001,Lima2005,Allen2011}. Clusters can be identified and measured by using cluster richness and weak lensing in photometric and spectroscopic surveys \citep{Rozo2010,Oguri2011,Oguri2014,Rozo2015,Rykoff2014,Simet2017}. The cluster catalog constructed from optical wavelengths and applied to constraining the cosmological parameters have been extensively studied recently \citep[e.g.][]{Haan2016,Sartoris2016,Heneka2018,Salvati2020,Fumagalli2021,Costanzi2019,Costanzi2021,Abdullah2020,Sunayama2020,Park2021}.

Ongoing surveys, such as DES, KiDS, HSC, and Extended Baryon Oscillation Spectroscopic Survey (eBOSS\footnote{https://www.sdss.org/surveys/eboss}), 
have played crucial roles in testing the $\Lambda$CDM model and constraining the cosmological  parameters. In the next decade, upcoming galaxy surveys such as Vera C. Rubin Observatory \citep[or LSST,][]{Chisari2019,Ivezic19}, Nancy Grace Roman Space Telescope (RST) \citep[or WFIRST,][]{Akeson19}, Euclid \citep{Laureijs11,Amendola2013,Amendola2018}, and CSST \citep{Zhan2011SSPMA,Zhan2018,Zhan2021,Gong19}, 
will explore an unprecedented large volume of the Universe, enabling us to test the cosmological model and probe the cosmological parameters with extraordinary precision. 

The CSST is a $2\,{\rm m}$ space telescope, which will cover 17500 deg$^2$ survey area with multi-band photometric imaging and slitless grating spectroscopy. It can perform photometric and spectroscopic surveys simultaneously with high spatial and spectral resolutions. It has $NUV$, $u$, $g$, $r$, $i$, $z$, and $y$ seven photometric imaging bands, which covers the wavelength range from 250\,nm to 1000\,nm. The 5$\sigma$ point-source magnitude limits for the seven bands can achieve 25.4, 25.4, 26.3, 26.0, 25.9, 25.2, and 24.4 AB mag, respectively. The CSST spectroscopic observation is accomplished by slitless gratings, which contains $GU$, $GV$, and $GI$ bands with the same wavelengths range and survey area as photometric observation. The corresponding magnitude limits are 23.2, 23.4, and 23.2, respectively. Since the CSST photometric and spectroscopic surveys will cover large and overlapping regions of the sky, they allow us to measure the growth and geometry of the Universe through a variety of cosmological probes, e.g. weak lensing \citep{Troxel2018,Secco2022}, photometric galaxy clustering \citep{Elvin2018,Porredon2021}, cluster number counts \citep{Haiman2001,Lima2005,Allen2011}, Alcock-Paczynski (AP) effect \citep{Alcock1979,Li2014ApJ,Li2019ApJ}, redshift-space distortions \citep[RSD,][]{GilMarin2016a,GilMarin2017a}, BAO \citep{GilMarin2016b,Beutler2017,Rosell2022,DES2021bao,Neveux2020,GilMarin2020}, etc. The CSST will be a powerful survey instrument for probing expansion history and structure growth of the Universe. Especially, Combining various CSST cosmological probes will provide more robust and precise constraints on cosmological parameters.

In this paper we forecast the cosmological constraints from the CSST 3$\times$2pt, spectroscopic galaxy clustering, and cluster number counts surveys. The mock data are generated based on the CSST observational and instrumental designs, and related systematics are also considered. We use Fisher matrix to extract cosmological information and perform the prediction. The analysis of the CSST photometric 3$\times$2pt and spectroscopic galaxy clustering surveys are discussed in section \ref{photoz} and \ref{specz}, respectively. The number counts of galaxy clusters for the CSST is shown in section \ref{cluster}. The details of Fisher matrix analysis is given in section \ref{fisher}. We present our constraint results of cosmological and systematical parameters in section \ref{discussion}, and summarize the results in section \ref{conclusion}.

\section{3$\times$2pt analysis}\label{photoz}

The CSST is expected to obtain billions of high-resolution galaxy images, that enables us to study the weak gravitational lensing effect and measure the evolution and formation of the LSS. Usually, the weak lensing effect is much smaller than the intrinsic irregularity of shapes and sizes of galaxies, and can not be detected in any single galaxy image. So we have to rely on statistical approaches to measure this signal (the so-called shear signal) from large galaxy samples, and extract the cosmological information encoded therein \citep{Munshi2008,Hoekstra2008,Kilbinger15,Mandelbaum2018}.

Besides the shape related information, the photometric observation also provides us with the spatial distribution information of galaxies (i.e. galaxy clustering).  This enables us to perform the galaxy clustering analysis using the angular power spectra in tomographic photometric redshift bins  \citep{Weaverdyck2021,Monroy2022}. In case of the uncertainties of photo-$z$ are well controlled, high precision cosmology results can be obtained \citep{Tanoglidis2020,Hasan2022} 

Furthermore, the cross-correlation between the lensing effect of sources and spatial distribution of lenses, known as the galaxy-galaxy lensing, \citep{Baldauf2010,Heymans2013,Mandelbaum2013,Singh2017}, severs as another complementary statistical method for the cosmological analysis. The cross-correlation approach has great advantages in mitigating critical sources of systematic errors, in particular the intrinsic alignment effects in the observed shear signal and galaxy bias \citep{Mandelbaum2005,Clampitt2017,Giblin2021,Park2021a}.

A combination of the above three methods leads to the so-called 3$\times$2pt method, which is widely applied in the analysis of ongoing observations \citep{Uitert2018MNRAS,Heymans21,Joachimi21,Abbott2022} and forecasts of near-future photometric surveys \citep{Euclid20,Tutusaus2020,Zuntz2021}. 
This combination approach is beneficial to breaking the degeneracy between cosmological parameters and galaxy bias, and also enables the self-calibration of the systematical and astrophysical parameters in a specific cosmological model. Therefore, in this work, we firstly study the application of the 3$\times$2pt method to the CSST photometric observation. 

\begin{figure}
	\centering
	\includegraphics[width=0.47\textwidth]{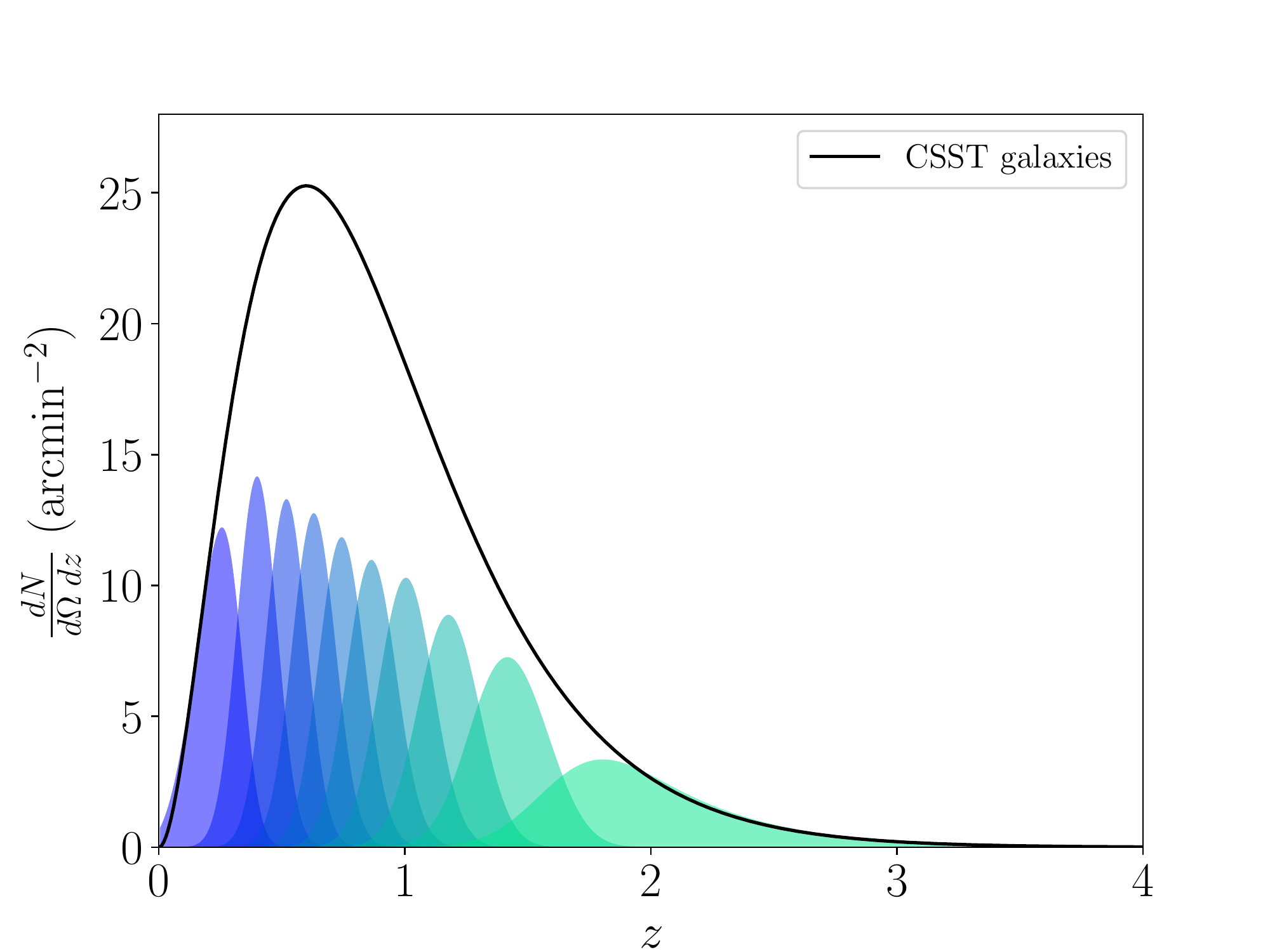}
	\caption{The galaxy redshift distribution of the CSST photometric survey. The total and 10 tomographic galaxy distributions are shown in black solid curve and filled regions, respectively.}
	\label{dndz}
\end{figure}

Following the CSST specifications given in \cite{Gong19}, we adopt the galaxy redshift distribution $n(z) \propto z^{\alpha} e^{-(z/z_0)^{\beta}}$, with $\alpha = 2$, $ \beta = 1 $, $ z_0 = 0.3 $, and choose a total number density $ n = 28\,\rm{arcmin^{-2}} $. We assume the same galaxy distribution for lenses and sources, since it can help to constrain the photo-$z$ error \citep{Schaan20}. To carry out a tomographic analysis, we divide the galaxy sample into 10 photo-$z$ bins with identical galaxy number density per bin. The redshift distribution of the galaxies as well as our binning scheme is illustrated in Fig. \ref{dndz}
\footnote{Here we stress that the cosmological results are dependent on the bin width \citep{Hu1999ApJ,zhan06,Asorey2012,Tanoglidis2020,Hasan2022}, and increasing the number of bins is always beneficial unless the bin widths are smaller than the individual photo-$z$ scatter. In addition, the scheme of an equal number of galaxies per bin can capture more information than equally spaced bins in redshift \citep[see e.g.][]{Taylor2018}. We will discuss this complicated issue in future works.}. 

Photo-$z$ uncertainty is one of the leading systematics in photometric observations \citep{Hildebrandt2021,stolzner2021,Rau2022}. For modeling of the photo-$z$ uncertainty, we assume that the true redshift distribution of a photometric tomographic bin is given by the conditional probability distribution $p\left(z_{\mathrm{ph}}|z\right)$ at a given $z$
\citep{Ma2006}, whose form in the $i$-th tomographic bin is described by 
\be
p^{i}\left(z_{\mathrm{ph}}|z,x\right)=\frac{1}{\sqrt{2\pi}\sigma_{z}}\exp\left[-\frac{\left(z-z_{\mathrm{ph}}-\delta^{i}_{z,x}\right)^{2}}{2\sigma_{z}^{2}}\right]\,,
\ee
where $x\in\{g,\kappa\}$ ($g$ and $\kappa$ denote galaxy clustering and weak lensing surveys, respectively), $\delta^{i}_{z,x}$ is the photo-$z$ bias, and the scatter parameter $\sigma_{z}$ is characterized as
\be
\sigma_{z} = \sigma_{z_{0}}(1+z)\,.
\ee
So the true redshift distribution of galaxies in a tomographic bin is given by
\be
n^{i}_x(z)=\int_{z^{i}_{\mathrm{min}}}^{z^{i}_{\mathrm{max}}}dz_{\mathrm{ph}}\,n_x(z)\,p^{i}\left(z_{\mathrm{ph}}|z\right)\,,
\ee
and the total surface number density of galaxies in a tomographic bin is 
\be
\bar{n}_{x}^{i}=\int dz\;n_{x}^{i}(z)\,.
\ee

The angular power spectrum $C_{AB}^{ij}(\ell)$ is calculated under the flat sky assumption and Limber approximation \citep{Kaiser1993} between tomographic bin $i$ of observable A and tomographic bin $j$ of observable B, where A, B $\in\{g, \kappa\}$. It can be written as
\be
C_{AB}^{ij}(\ell)=\int d\chi\frac{q_{A}^{i}(\chi)q_{B}^{j}(\chi)}{\chi^{2}}P_{m}\bigg(\frac{\ell+1/2}{\chi},z(\chi)\bigg),
\ee
where $\chi$ is the comoving radial distance, $P_m$ is the non-linear matter power spectrum, and $q_{A}^{i}$ are weight functions of the different observables. To constrain cosmological parameters with future high-precision data sets, the matter power spectrum must be modeled accurately. It has been found that the cosmic shear power spectrum is sensitive to a small scale until the wavenumber $k\simeq7\,h/{\rm Mpc}$ \citep{Taylor2018}. Our non-linear matter power spectrum $P_m$ is calculated using the $\tt halofit$ code \citep{Takahashi2012}.\footnote{Here we do not consider the baryonic effects on the non-linear power spectrum, which are important and have been studied extensively \citep{Mead2015,Mead2016,Mead2021,Copeland2018,Schneider2019,Huang2019,Martinelli2021}. We should note that if baryonic effects are considered, the results of weak lensing constraint on cosmological parameters may be degraded based on Fisher analysis \citep[see e.g.][]{Copeland2018}. We will study the baryonic effects particularly in our future work.} For the galaxy clustering and galaxy-galaxy lensing angular power spectrum, we only consider the multipoles that satisfy $\ell \leq k_{\rm max}\,\chi(z_{i}) -1/2$, where $z_i$ represents the mean value of redshift bin $i$. We have set $k_{\rm max} = 0.3\,h/{\rm Mpc}$ in the CSST photometric galaxy clustering and galaxy-galaxy lensing surveys to avoid the non-linear effects \citep{Schaan20}.

The weight function of the galaxy clustering and weak lensing are given by
\be 
q_{\mathrm{g}}^{i}(k,\chi)=b^{i}\left(k,z(\chi)\right)\frac{n_{\mathrm{g}}^{i}(z(\chi))}{\bar{n}_{\mathrm{g}}^{i}}\frac{dz}{d\chi}\,,
\ee

\be
q_{\kappa}^{i}(\chi)=\frac{3H_{0}^{2}\Omega_\mathrm{m}}{2c^{2}}\frac{\chi}{a(\chi)}\int_{\chi}^{\chi_{\mathrm{h}}}\mathrm{d}\chi^{\prime}\frac{n_{\kappa}^{i}(z(\chi^{\prime}))dz/d\chi^{\prime}}{\bar{n}_{\kappa}^{i}}\frac{\chi^{\prime}-\chi}{\chi^{\prime}}\,,
\ee
where $b_i$ is the linear galaxy bias in the $i$th bin, $c$ is the speed of light, $a$ is the scale factor, $H_0$ is the Hubble constant, and $\chi_{\mathrm{h}}$ is the line-of-sight comoving horizon distance.

The shear signal is coherent with intrinsic alignment (IA) effect. If galaxy shapes are randomly oriented, IA will not affect the results of shear analysis. However, because of the local tidal fields and gravitational interactions between galaxies, IA leads to an additional term for the shear signal. Therefore the IA must be modeled in our observed shear signal. The exact form of IA is related to galaxy formation and evolution, and also halo models \citep{Joachimi2015,Kiessling2015,Kirk2015,Troxel2015}. Due to the uncertainty of those processes, IA is typically predicted by linking observed galaxy shapes to the gravitational tidal field in the LSS based on tidal alignment or tidal torquing models \citep{Hirata2004PhRvD,Bridle2007NJPh,Blazek2011JCAP}. We assume the tidal alignment model in our analysis, which describes the shape alignments of red elliptical galaxies in the tidal field. It captures the primary IA effect, and ignores the tidal torquing mechanism and secondary alignment of blue spiral galaxies \citep{Blazek2019PhRvD,Samuroff2019MNRAS}.
When IA is considered, the galaxy lensing weight function can be modeled as \citep{Kirk2012,Krause2016} 
\be
q_{\kappa}^{i}(\chi) \longrightarrow q_{\kappa}^{i}(\chi)-A\left(z\left(\chi\right)\right)\frac{n^{i}_{\kappa}(z(\chi))}{\bar{n}^{i}_{\kappa}}\frac{dz}{d\chi}\,.
\ee
Then the final shear power spectrum will contain the auto-correlation of IA between neighboring galaxies, and the cross-correlation between the IA of foreground lens and shear signals of background sources.
The function $A\left(z\left(\chi\right)\right)$ can be expressed as
\be
A(z)=A_{{\rm IA}}C_{1}\rho_{\rm cr}\frac{\Omega_\mathrm{m}}{D(z)}\left(\frac{1+z}{1+z_{0}}\right)^{\alpha_{{\rm IA}}}\left(\frac{L}{L_{0}}\right)^{\beta_{{\rm IA}}}\,,
\ee
where $A_{\rm IA}$ is intrinsic alignments amplitude, $\rho_{\rm cr}$ is the present critical density and  we have $C_{1}\rho_{\rm cr} = 0.0134$, $z_{0} =  0.62$ is the pivot redshift, $D(z)$ is the growth factor, $\alpha_{\mathrm{IA}}$ and $\beta_{{\rm IA}}$ represent the relations of redshift and luminosity, respectively. For simplicity, we fix $\beta_{{\rm IA}} = 0$ for ignoring the dependence of luminosity.

In addition to IA, the process of estimating shear signals from galaxy shapes also leads to additional biases, which are commonly described as the additive and multiplicative errors \citep{Huterer2006,Heymans2006,Massey2013MNRAS}. To satisfy the requirements of future weak lensing surveys, the additive and multiplicative errors need to be smaller than $10^{-3}$ and 0.03 at least, respectively. In particular, the multiplicative bias degenerates with the shear amplitude, and can significantly affect the accuracy of shear measurements. Many uncertainties can result in multiplicative bias \citep{Jarvis2016MNRAS,Fenech2017MNRAS}, such as errors from point spread function (PSF) modeling size, select bias, galaxy shape measurement, or detector systematics \citep{Paulin2008,Mandelbaum2018,Pujol2020A}. Since the upcoming Stage IV weak lensing survey will measure the shear signal with an unprecedented level of accuracy, shear biases must be calibrated precisely \citep{Taylor2018MNRAS,Gillis2019MNRAS}. For example, the multiplicative bias can be self-calibrated by cross-correlation with CMB lensing \citep{Vallinotto2012ApJ,Das2013arXiv1311,Schaan17}, and machine learning method also can be used in the calibration \citep{Pujol2020b}. We consider the multiplicative shear calibration in our analysis and describe it as one parameter $m^i$ in tomographic redshift bin $i$. For the weak lensing survey and galaxy-galaxy lensing survey, we have

\begin{equation*}
C_{\kappa\kappa}^{ij}(\ell)\quad \longrightarrow \quad(1+m^{i})\,(1+m^{j})\,C_{\kappa\kappa}^{ij}(\ell)\,,
\end{equation*}
\be
C_{\mathrm{g}\kappa}^{ij}(\ell)\quad \longrightarrow \quad(1+m^{j})\,C_{\delta_{\mathrm{g}}\kappa}^{ij}(\ell)\,.
\ee

For galaxy clustering survey, the linear galaxy bias is modeled using one nuisance parameter in a tomographic redshift bin. The fiducial values of galaxy bias is given by $b(z) = 1 + 0.84z$ \citep{zhan06}.
Besides, since the measured power spectra also subject to galaxy shot noise and other systematic noises, and then we have
\be
\tilde{C}_{\kappa\kappa}^{ij}(\ell) = C_{\kappa\kappa}^{ij}(\ell) +\delta^{\rm K}_{ij}\frac{\sigma_{\epsilon}^2}{\bar{n}_{\kappa}^{i}} + N_{\kappa\kappa}\,,
\ee
\be
\tilde{C}_{gg}^{ij}(\ell) = C_{gg}^{ij}(\ell) +\delta^{\rm K}_{ij}\frac{1}{\bar{n}_{\rm g}^{i}} + N_{gg}\,,
\ee
where $\delta^{\rm K}_{ij}$ is Kronecker delta function, $\sigma_{\epsilon}^2$ is the variance of the observed ellipticities, $N_{\kappa\kappa}$ and $N_{gg}$ are systematic noises that may be generated from the PSF, photometry offsets, instrumentation effects, dust extinction, and so on \citep{Tegmark2002,Jain2006}. Here, we just simply assume $N_{\kappa\kappa} = 10^{-9}$ and $N_{gg} = 10^{-8}$ \citep{zhan06,Huterer2006,Gong19}, and $\sigma_{\epsilon} = 0.3$ in our analysis.

\section{Spectroscopic galaxy clustering}\label{specz}

\begin{figure}
	\centering
	\includegraphics[width=0.47\textwidth]{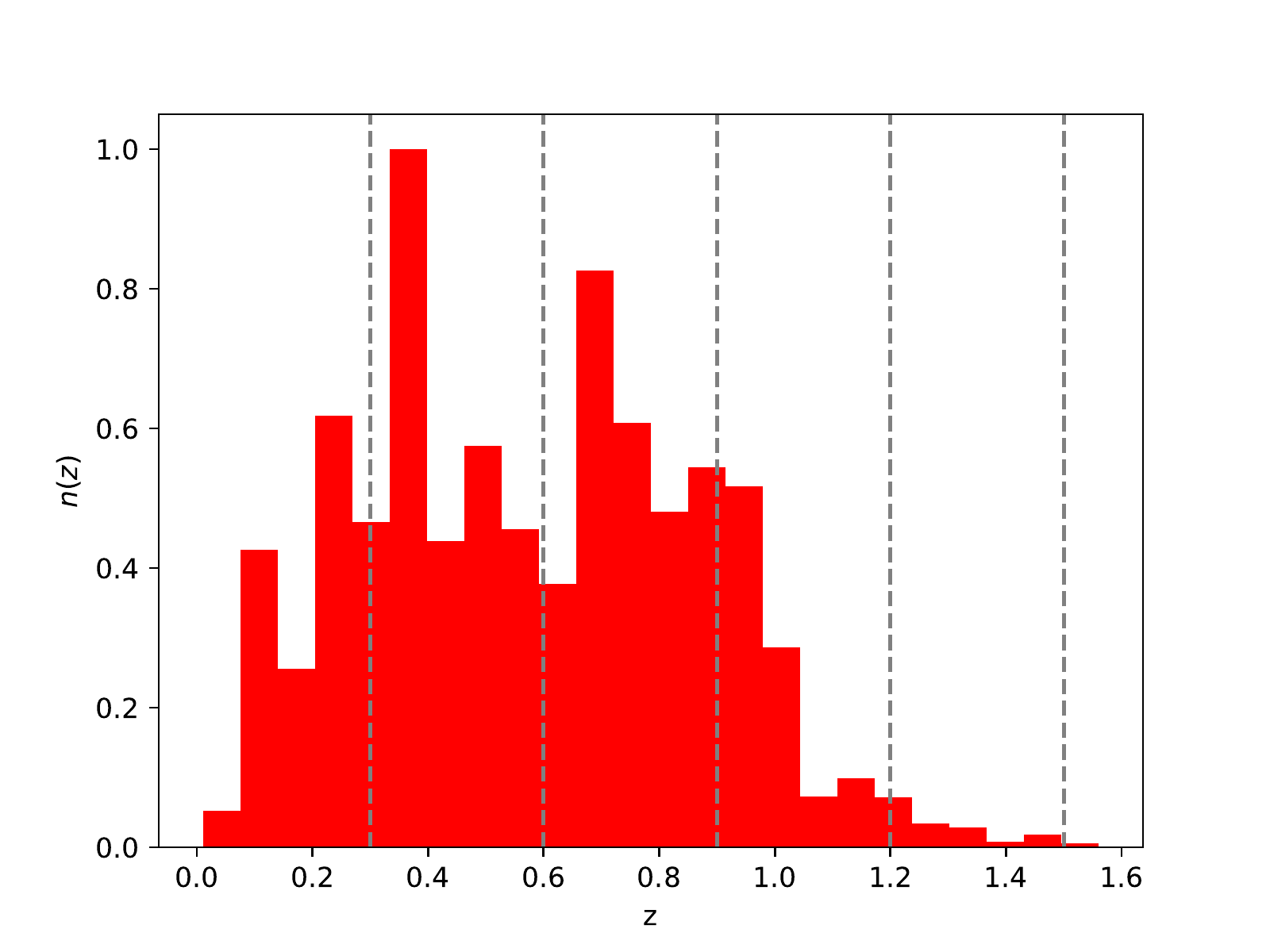}
	\caption{Galaxy number density distribution of the CSST slitless spectroscopic survey. The vertical grey dashed lines denote the five spec-$z$ bins we divide for cosmic evolution study.}
	\label{spec-z}
\end{figure}

Besides photometric imaging survey, the CSST can also perform slitless spectroscopic galaxy survey simultaneously, and more than one hundred million galaxy spectra (mainly from emission line galaxies) can be obtained \citep{Gong19}. 
Based on the zCOSMOS catalog \citep{Lilly2009}, we simulate the mock catalog of CSST spectroscopic galaxy survey, and divide the redshift range from $z=0$ to 1.5 into five bins for studying evolution of important cosmological parameters, such as the equation of state of dark energy. The details can be found in \cite{Gong19}.  The expected galaxy number density distribution of the CSST spectroscopic survey has been shown in Fig. \ref{spec-z}, and the details of surface and volume number densities and galaxy bias in each spec-$z$ bins are given in Table \ref{spec}. 

From the current spectroscopic galaxy surveys, such as the 2-degree Field Galaxy Redshift Survey \citep[2dFGRS,][]{Cole2005MNRAS} to 6-degree Field Galaxy Survey  \citep[6dFGS,][]{Jones2009MNRAS}, WiggleZ Dark Energy Survey \citep[WiggleZ,][]{Parkinson2012PhRvD}
 then to SDSS-III Brayon Oscillation Spectroscopic Survey \citep[BOSS,][]{Dawson2013AJ} and SDSS-IV extended BOSS \citep[eBOSS,][]{Dawson2016AJ}, we can acquire a wealth of information to understand our Universe. The BAO and RSD signals from these surveys have been analyzed particularly  by using different methods \citep{Beutler2017,Beutler2017b,GilMarin2020,Foroozan2021JCAP}. 
One can obtain important information of the LSS and dark energy from the BAO wiggles of the galaxy power spectrum \citep{Seo2007,Ribera2014}, or the full shape of the power spectrum which contains more information. 

\begin{table}
	\caption{Galaxy surface and volume number densities, and galaxy biases for the five spec-$z$ bins in the CSST spectroscopic survey.}
	\begin{center}
	\scalebox{0.8}{
		\begin{tabular}{c c c c c c}
			\hline
			\hline
			$z_{\rm min}$ & $z_{\rm max}$ & $z_{\rm mean}$ & d$N(z)$/d$\Omega$d$z{\rm[arcmin^{-2}]}$ & $\bar{n}(z)[h^3{\rm Mpc^{-3}}]$ & $b(z)$  \\
			\hline
			0 & 0.3 & 0.15 & 1.54 & 2.81 $\times 10^{-2}$ & 1.126 \\
			0.3 & 0.6 & 0.45 & 3.35 & 1.16 $\times 10^{-2}$ & 1.378 \\
			0.6 & 0.9 & 0.75 & 3.16 & 5.63 $\times 10^{-3}$ & 1.63 \\
			0.9 & 1.2 & 1.05 & 0.9 & 1.15 $\times 10^{-3}$ & 1.882 \\
			1.2 & 1.5 & 1.35 & 0.09 & 9.65 $\times 10^{-5}$ & 2.134 \\
			\hline
		\end{tabular}
		}
	\end{center}
	\label{spec}
\end{table}

In a real spectroscopic survey, we are actually observing the distribution of galaxies in redshift space. The observed galaxy redshift always contains an additional contribution by the peculiar velocity of the galaxy, which can be expressed by
\be
1 + z_{\rm obs} = (1 + z)(1 + v_{||}/c)\,,
\ee
where $v_{||}$ denotes the peculiar velocity of a galaxy along the light of sight. This term will lead to the redshift-space clustering anisotropic of galaxy distribution, or so-called RSD \citep{Hamilton1998ASSL}. The relation between real and redshift space galaxy power spectrum has been given in linear regime \citep{Kaiser1987MNRAS}. When analyzing galaxy distribution from galaxy surveys, we must assume a fiducial cosmological model to transform observed angular positions and redshift into physical coordinates. If we adopt a wrong fiducial cosmology, an extra anisotropic clustering signal will be introduced. This distortion, which is called AP effect, is different from RSD and also can be used to constrain cosmological parameters \citep{Marinoni2010Nature,Lopez2014ApJ,Li2018ApJ,Li2019ApJ}. 
Here the linear galaxy bias $b(z)$, RSD, and AP effect are considered, and the galaxy power spectrum can be written as 
\be
\begin{split}
P_\text{obs}(k_\text{ref},\mu_\text{ref};z)=&\frac{1}{q_\perp^2 q_\parallel}\left(b(z)\sigma_8(z)+f\sigma_8(z)\mu^2\right)^2\frac{P_{\rm m}(k;z)}{\sigma_8^2(z)}\\
&\times e^{-k^2\mu^2\sigma_r^2(z)}+ P_\text{s}(z)\,,
\end{split}
\label{pklin}
\ee
where $f={\rm d\,ln}D(a)/{\rm d\,ln}a$ is the growth rate, $P_\text{s}(z)$ is the shot noise term, and $P_{\rm m}(k;z)$ is the matter power spectrum. Here only the linear matter power spectrum is considered, which can be obtained by $\tt CAMB$ \citep{Lewis2000}. 
The radial smearing factor $\sigma_r(z)$ is induced by spec-z error and can be estimated as
\be
\sigma_r(z) = \frac{\partial{r}}{\partial{z}} \sigma_z(z) 
= \frac{c}{H(z)} \left(1+z\right) \sigma_{0,z} \, ,
\ee
where $\sigma_z(z)=\sigma_{0,z}(1+z)$, and $\sigma_{0,z}$ is the error of redshift measurement. 
The $k$ and $\mu$ of eq. (\ref{pklin}) are given by 
\bea
&k(k_\text{ref},\mu_\text{ref}) = \frac{k_\text{ref}}{q_\perp} \left[ 1 + \mu_\text{ref}^2 \left(\frac{q_\perp^2}{q_\parallel^2} - 1 \right) \right]^{1/2}\,, \nonumber\\
&\mu(\mu_\text{ref}) = \mu_\text{ref} \frac{q_\perp}{q_\parallel} \left[ 1 + \mu_\text{ref}^2 \left(\frac{q_\perp^2}{q_\parallel^2} - 1 \right) \right]^{-1/2} \, . 
\eea
where 
\bea
&q_{\perp}(z) = \frac{D_{\rm A}(z)}{D_{\rm A,\, ref}(z)}\,,
\qquad   q_{\parallel}(z) = \frac{H_\text{ref}(z)}{H(z)} \,,\nonumber\\
&k_{\perp} = \frac{k_{\perp,\text{ref}}}{q_\perp}
\qquad \text{and} \qquad
k_{\parallel} = \frac{k_{\parallel,\text{ref}}}{q_\parallel} \,.
\eea
The indicator `ref' denotes the referenced cosmology, and $D_{\rm A}(z)$ and $H(z)$ represent the angular diameter distance and Hubble parameter, respectively. For the galaxy power spectrum, we do not consider non-linear effects, since we are focusing on linear scales here. Galaxy bias is modeled as $b(z) = 1+ 0.84z$, the redshift uncertainty is assumed to be $\sigma_{0,z} = 0.002$ for the CSST, and shot noise $P_\text{s}(z) = 1/\bar{n}$. Besides, we also add a constant systematics term, $N_{\rm{sys}} = 5\times10^4\,({\rm Mpc}/h)^3$, in galaxy power spectrum to consider instrumentation effects \citep{Gong19}.

\section{Cluster number counts}\label{cluster}

Galaxy cluster abundance or number counts is another important probe in the CSST survey. 
Clusters can be detected at different bands with various methods, including X-ray, optical, and sub-mm bands \citep{Koester2007ApJ,Rozo2015,Sunyaev1972CoASP}, and cluster abundance is a powerful tool in cosmological studies. However, accurately identifying cluster members and determining cluster mass are still challenging in current galaxy surveys, which are crucial in the studies of number counts of galaxy clusters. 
A class of methods for finding galaxy clusters and estimating their mass in optical surveys is based on the presence of a red sequence of galaxies, such as maxBCG \citep{Koester2007ApJ}, redMaPPer \citep{Rykoff2014} used in DES \citep{Rykoff2016ApJS}, and CAMIRA \citep{Oguri2014} applied in HSC \citep{Oguri2018PASJ}. Besides, a different approach, e.g. AMICO \citep{Bellagamba2018MNRAS}, which is based on the optimal filtering technique, has been used in KiDS  \citep{Bellagamba2019MNRAS,Lesci2022}. Usually, the cluster mass can be measured using observational proxies that can be related to mass. In optical surveys, cluster richness and gravitational lensing effect often serve as mass tracers \citep{Rozo2009ApJ,Rykoff2012ApJ}. Based on the mass-observable relations or so-called scaling relations, one can estimate the cluster mass in an individual or statistical way \citep{Rozo2010,Abdullah2020,Costanzi2021,Lesci2022A&A}.

\begin{figure}
	\centering
	\includegraphics[width=0.47\textwidth]{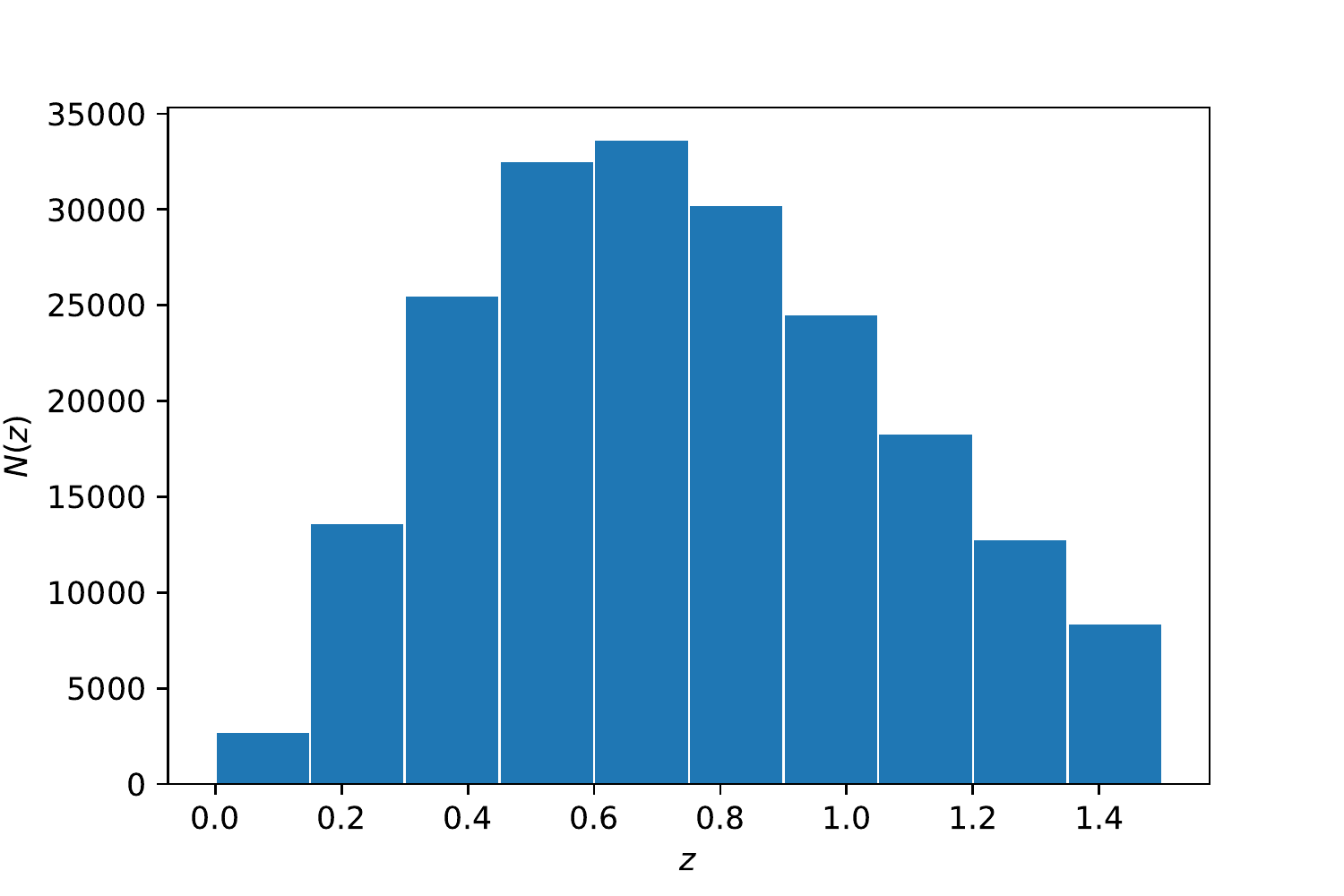}
	\caption{The expected redshift distribution of number counts of galaxy clusters in the CSST survey. The redshift range we consider from $z=0$ to 1.5 has been divided into 10 bins.}
	\label{number}
\end{figure}

The CSST can perform photometric and spectroscopic surveys simultaneously, which is powerful to identify galaxy clusters and their members, and measure their redshifts and masses. Here we derive the CSST number counts of galaxy clusters from the halo mass function, which is given by \citep{Press1974,Sheth1999,Tinker2008}
\be
\frac{{\rm d}n(M,z)}{{\rm d}\ln M}=\frac{\bar{\rho}_{\rm m}}{M}\,f(\sigma)\,\frac{{\rm d}\ln\sigma^{-1}}{{\rm d}\ln M}\,,
\ee
where $\bar{\rho}_{\rm m}$ is the mean matter density, and  
\be 
\sigma^{2}(R)=\frac{1}{2\pi^{2}}\int{\rm d}k\;k^{2}\,P(k)\,W^{2}_{R}(k)\,.
\ee
Here $R = (3M/4\pi\bar{\rho}_{\rm m})^{1/3}$ is the smoothing scale, $W(k,R)=\frac{3}{(kR)^{3}}(\sin kR-(kR)\cos kR)$ is the window function. 
The multiplicity function is given by \citep{Tinker2008}
\be
f\left(\sigma\right)=A\left[\left(\frac{\sigma}{b}\right)^{-a}+1\right]e^{-c/\sigma^{2}}\,,
\ee
where the parameters A, a, b, and c depend on the redshift and halo mass definition and can be determined by simulations. 

The expected values of cluster number counts in the $i$-th mass bin and $\alpha$-th redshift bin is then given by
\be
N_{\alpha i}=\int_{\Delta z_{\alpha}}{\rm d}z\;\frac{{\rm d}V}{{\rm d}z}\int_{\Delta M_{i}}{
\rm d}M\;\frac{{\rm d}n}{{\rm d}M}(M,z)\,.
\ee
Here,
\be
\frac{dV}{dz}=\Delta\Omega\left(\frac{\pi}{180}\right)^{2}\frac{c}{H(z)}\left(\int_{0}^{z}dz^{\prime}\frac{c}{H(z^{\prime})}\right)^{2}\,,
\ee
where $\Delta\Omega$  is the survey area in square degrees. We use the public CCL code \citep{Chisari2019} to calculate the cluster abundance. For the CSST, to accurately measure the redshifts and masses of galaxy clusters, we only consider the redshift range from $z=0$ to 1.5, which can be covered by both of CSST photometric and spectroscopic surveys. This redshift range is equally divided by 10 bins to study the evolution of the Universe. We consider the cluster halo mass range from 10$^{14}$\,$h^{-1}M_{\odot}$ to 10$^{16}$\,$h^{-1}M_{\odot}$ in the estimation. We find that the CSST could detect about 170,000 clusters in 17500\,deg$^2$ survey area. The redshift distribution of the number counts of CSST galaxy clusters has beem shown in Fig. \ref{number}. We find that the CSST cluster redshift distribution has a peak around $z=0.6$, and can extend to $z\sim2$.

Besides the statistical uncertainties including shot noise and sample variance, the uncertainties of cluster number counts mainly come from observational uncertainties related to mass measurements, such as photometric/spectroscopic noise, background subtraction, and projection/percolation effects \citep[see e.g.][]{Costanzi2019}. We will discuss these uncertainties in details in the next section.

\section{Fisher matrix forecast}\label{fisher}

We adopt Fisher matrix to predict the constraint results from the cosmological multi-probes in the CSST photometric and spectroscopic surveys. We assume the Likelihood function to be Gaussian and is given by 
\be
\ln L({\vec D}| {\vec \theta})=-\frac{1}{2}\left({\vec D}-{\vec M}({\vec \theta})\right)^{T}\Sigma^{-1}\left({\vec D}-{\vec M}({\vec \theta})\right),
\ee 
where $\vec D$ is the data vector, $\vec M({\vec \theta})$ is the corresponding predicted vector from theoretical model given the parameter vector $\vec \theta$, and $\Sigma$ is the covariance matrix. Then the Fisher matrix is defined as \citep{Tegmark1997}
\be
F_{\theta_{\alpha}\theta_{\beta}} = \bigg\langle-\frac{\partial^2\ln L}{\partial\theta_{\alpha}\partial\theta_{\beta}}\bigg|_{\theta_{\rm{0}}}\bigg\rangle\,,
\ee
where the derivatives are calculated in the point $\theta_{\rm{0}}$ of the parameter space. If assuming the covariance matrix is independent of the parameters, which should be available for most cases \citep{Carron2013}, it can be rewritten as
\be
F_{\theta_{\alpha}\theta_{\beta}}=\frac{\partial M^{T}}{\partial\theta_{\alpha}}\Sigma^{-1}\frac{\partial M}{\partial\theta_{\beta}}.
\ee 
Besides, we can easily get a new Fisher matrix for a set of new parameters $\vec p (\vec \theta)$ by performing Jacobian transform
\be \label{eq:Jacobian_fisher}
S_{p_{i}p_{j}} = \sum_{\alpha\beta} \frac{\partial \theta_{\alpha}}{\partial p_{i}}{F}_{\theta_{\alpha}\theta_{\beta}}\frac{\partial \theta_{\beta}}{\partial p_{j}} \, .
\ee

In our forecast, the final adopted cosmological parameters for the three CSST cosmological probes are
\be
\vec \theta = \{\Omega_{\rm m},\,\Omega_{\rm b},\,w_0,\, w_{\rm a},\,h,\, n_{\rm s},\,\sigma_8\},
\ee
and the fiducial values of those cosmological parameters are assumed to be $(0.314\,, 0.0494\,, -1\,, 0\,, 0.6732\,, 0.966\,, 0.82)$.

\subsection{Forecast for photometric 3$\times$2pt surveys}
In the case of 3 $\times$ 2pt analysis, the covariance matrix can be expressed as a four point function, and it can be estimated as 
\bea
{\rm Cov}_{ABA^\prime B^\prime}^{(ijkl)}(\ell)\equiv{\rm Cov}\left[\tilde{C}^{ij}_{AB}(\ell),\tilde{C}_{A^\prime B^\prime}^{kl}(\ell^\prime)\right] \nonumber \\ =\frac{\tilde{C}_{AA^\prime}^{ik}(\ell)\tilde{C}_{BB^\prime}^{jl}(\ell^\prime)+\tilde{C}_{AB^\prime}^{il}(\ell)\tilde{C}_{BA^\prime}^{jk}(\ell^\prime)}{(2 \ell + 1) f_{\rm sky} \Delta \ell}\delta^{\rm K}_{\ell\ell^\prime}. \label{eq: covfourth}
\eea
Here $f_{\rm sky}$ is the sky survey fraction, which is about $42\%$ for the CSST survey. The data vector of the observed weak lensing, galaxy-galaxy lensing, and angular galaxy power spectra $D^{\rm T}$ is given by 
\begin{align}
D^{\rm T} =&\biggl{\{}\tilde{C}^{(11)}_{\kappa\kappa}(\ell),\,..\,,\tilde{C}^{(N_{\rm zbin}N_{\rm zbin})}_{\kappa\kappa}(\ell), \tilde{C}^{(11)}_{g\kappa}(\ell),\tilde{C}^{(N_{\rm zbin}N_{\rm zbin})}_{g\kappa}(\ell), \nonumber \\
&\tilde{C}^{(11)}_{gg}(\ell),\,..\,,\tilde{C}^{(N_{\rm zbin}N_{\rm zbin})}_{gg}(\ell)\biggr{\}
}\,,
\end{align}
and the corresponding covariance is written as
\be
{\rm Cov}(\ell)=\left(\begin{array}[]{c|c|c}{\rm Cov}_{\kappa\kappa\kappa\kappa}^{(ijkl)}(\ell)&{\rm Cov}_{\kappa\kappa g\kappa}^{(ijkl
		)}(\ell)&{\rm Cov}_{\kappa\kappa gg}^{(ijkl)}(\ell)\\
	\hline\cr{\rm Cov}_{g\kappa\kappa\kappa}^{(ijkl)}(\ell)&{\rm Cov}_{g\kappa g\kappa}^{(ijkl)}(\ell)&{\rm Cov}_{g\kappa gg}^{(ijkl)}(\ell)\\
	\hline\cr{\rm Cov}_{gg\kappa\kappa}^{(ijkl)}(\ell)&{\rm Cov}_{ggg\kappa}
	^{(ijkl)}(\ell)&{\rm Cov}_{gggg}^{(ijkl)}(\ell)\end{array}\right)\;.
\ee
Then the final Fisher matrix is
\be
F_{\theta_{\alpha} \theta_{\beta}} = \sum_{m,n}^{N_{D}}\sum_{\ell}^{N^{\rm max}_{\ell}(m,n)}\frac{
	\partial{ D^{\rm T}}_{m}(\ell)}{\partial \theta_{\alpha}}\;{\rm Cov}^{-1}_{mn}(\ell)\;
\frac{\partial{D}_{n}(\ell)}{\partial \theta_{\beta}}\;,
\ee
where $N_D$ is the dimension of data vector, $N^{\rm max}_{\ell}$ is the available multipole mode bins.

\begin{table}
\caption{The $\ell$-cut in galaxy clustering angular power spectrum.}
\centering
\scalebox{1.}{
\begin{tabular}{c| c c c c c c c}
\hline
\hline
\multicolumn{1}{c}{$z_{\rm bin}$} & \multicolumn{1}{c}{$z_{\rm mean}$} & \multicolumn{1}{c}{$\ell_{\rm max}$} & \multicolumn{1}{c}{$N^{\rm max}_{\ell}$} \\
\hline
1 & 0.165 & 211  & 21  \\
2 & 0.395 & 476 & 30  \\
3 & 0.515 & 601 & 33 \\
4 & 0.625 & 707 & 34 \\
5 & 0.74 & 811  & 36  \\
6 & 0.865 & 916 & 37 \\
7 & 1.0 & 1025 & 38 \\
8 & 1.18 & 1149 & 40 \\
9 & 1.44 & 1312 & 41 \\
10 & 2.8 & 1894 & 45 \\
\hline
\end{tabular}
\label{cut}
}
\end{table}

The Fisher matrix analysis of the 3$\times$2pt is completed by using {\tt COSMOSIS} \citep{Zuntz2015}. To compute the Fisher matrix, we use $N_{\ell} = 50$ logarithm-spaced multipole bins between $\ell = 30$ and $\ell = 3000$. To avoid non-linear effect, we also remove the data at wavenumber $k_{\rm max}=0.3\ h/\rm Mpc$  in the photometric galaxy clustering and galaxy-galaxy lensing surveys. The corresponding $\ell_{\rm max}$ and $N^{\rm max}_{\ell}$ of the 10 redshift tomographic bins are listed in Table \ref{cut}. Besides the cosmological parameters, we also consider the systematic parameters in the 3$\times$2pt analysis, including the photo-$z$ bias, photo-$z$ scatter, intrinsic alignment, galaxy bias, and shear calibration errors. Their fiducial values are given by $\Delta^i_z = 0$, $\sigma_{z_0} \sim N({\rm mean}=0.05,\sigma=0.003)$ (Normal distribution) \citep{Ma2006,Schaan20}, $A_{\rm IA} = 1$, $\alpha_{IA} = 0$, $b_i = 1+0.84z_i$, and $m_i = 0$, respectively.

\begin{figure*}
        \centering
	\includegraphics[width=1.0\textwidth]{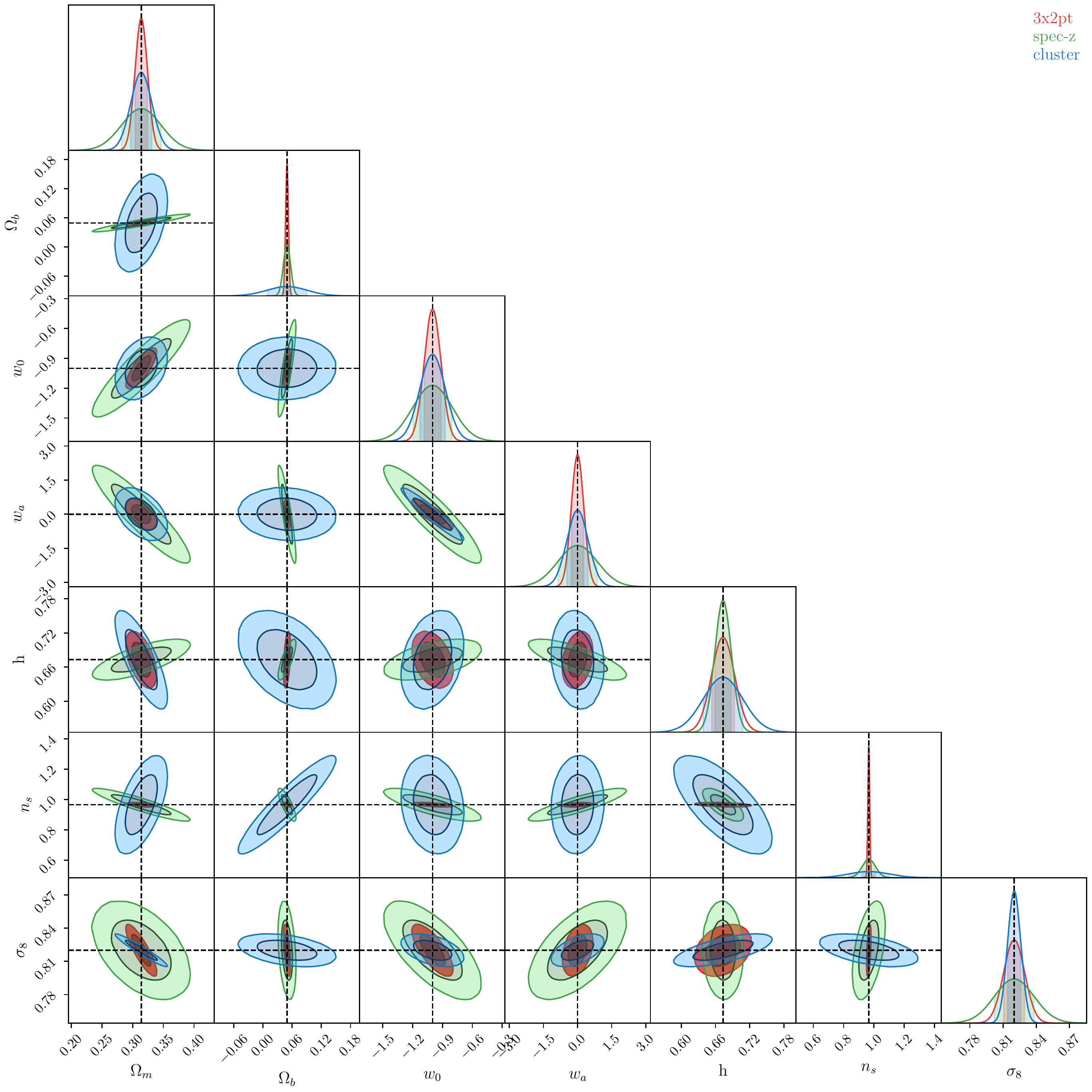}
    \caption{Marginalized posterior contour maps (1$\sigma$ and 2$\sigma$) and 1D PDFs of the seven cosmological parameters for the CSST 3$\times$2pt, spectroscopic galaxy clustering, and cluster number counts surveys, respectively.}
    \label{all}
\end{figure*}

\begin{figure*}
        \centering
	\includegraphics[width=1.0\textwidth]{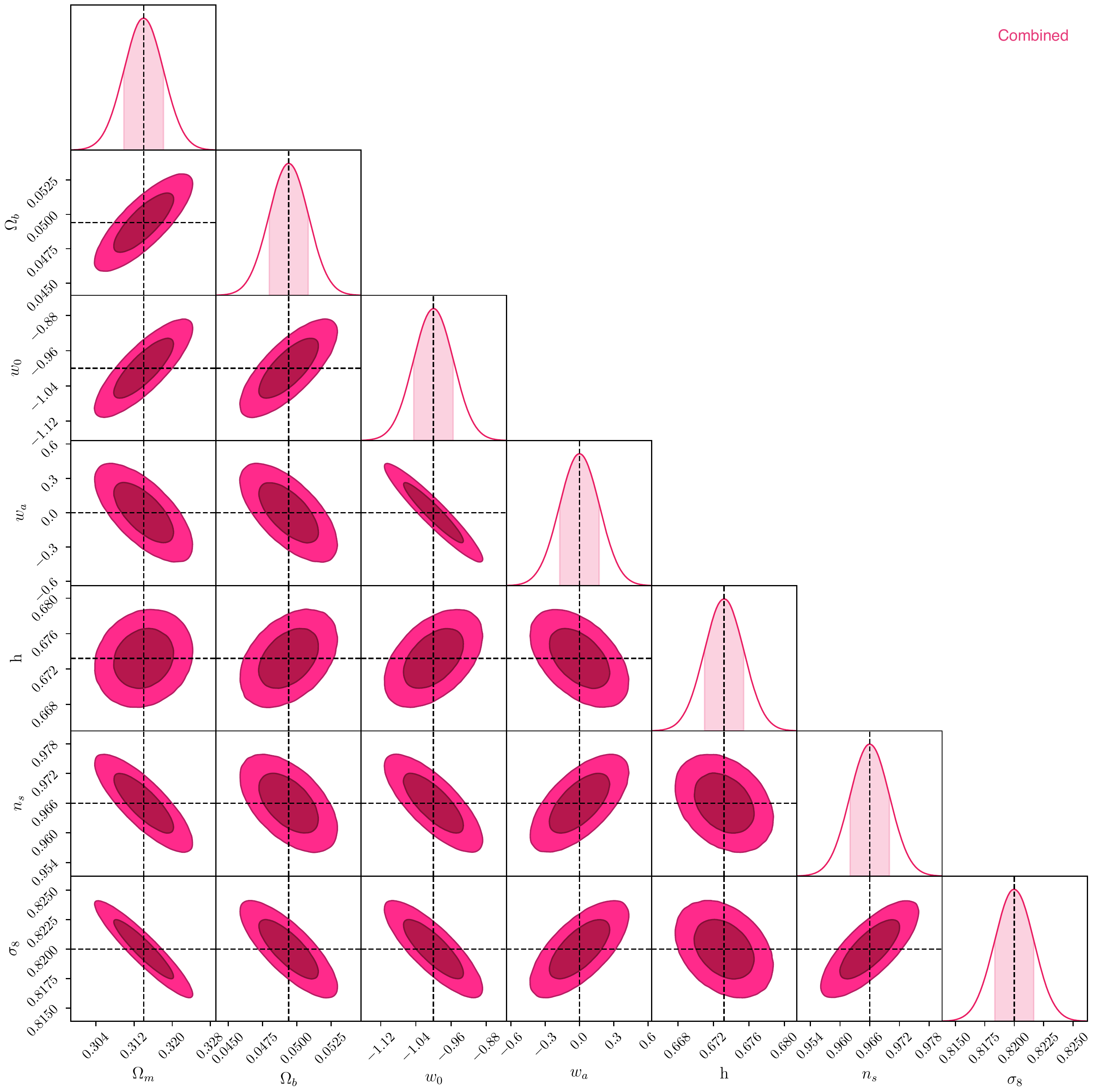}
    \caption{Marginalized posterior contour maps (1$\sigma$ and 2$\sigma$) and 1D PDFs of the seven cosmological parameters for the joint constraints of all three probes.}
    \label{combined}
\end{figure*}

\subsection{Forecast for spectroscopic galaxy clustering survey}\label{fisher_spec}
As listed in Table \ref{spec}, in the analysis of the CSST spectroscopic galaxy clustering survey, we divide the redshift range from $z = 0$ to $1.5$ into 5 tomographic redshift bins. In order to calculate Fisher matrix, we take the following cosmological parameters, i.e. shape parameters related to the shape of matter power spectrum $\{{\Omega_{\rm b}}, {\Omega_{\rm m}}, h, n_s\}$, and redshift-dependent parameters $\{\ln D_\text{A}(z_{i}), \ln H(z_{i}), \ln[f\sigma_8(z_{i})], \ln[b\sigma_8(z_{i})], P_\text{s}(z_{i})\}$ \citep{Wang2013MNRAS,Euclid20}. Note that the constraints on the other parameters, e.g. equation of state of dark energy $w_0$ and $w_a$, can be derived from the shape and redshift-dependent parameters by transferring the Fisher matrix into new ones using Eq.(\ref{eq:Jacobian_fisher}).

The Fisher matrix for spectroscopic galaxy surveys in a redshift bin is given by \cite{Tegmark1997PhRvL}
\bea
F_{\theta_{\alpha}\theta_{\beta}}^\textrm{bin}(z_{ i}) = \frac{1}{8\pi^2} 
\int^1_{-1} d\mu
\int^{k_\text{max}}_{k_\text{min}}{\rm d}k\;k^2V_\text{eff}(z_{ i};k,\mu)\nonumber \\
\left[\frac{\partial \ln P_\text{obs}(k,\mu ;z_{ i})}{\partial \theta_{\alpha}} 
\frac{\partial \ln P_\text{obs}(k,\mu ;z_{ i})}{\partial \theta_{\beta}}\right] 
\,,
\eea
where the effective survey volume is given by
\be 
V_\text{eff}(k,\mu;z) = V_\text{s} (z)\left[\frac{n(z) P_\text{obs}(k,\mu;z)}{n(z) P_\text{obs}(k,\mu;z) +1}\right]^2\,.
\ee
The total Fisher matrix of all redshift bins then can be estimated by
\be
F_{\theta_{\alpha}\theta_{\beta}} = \sum_{i=1}^{N_{z_\text{bin}}} F_{\theta_{\alpha}\theta_{\beta}}^\textrm{bin}(z_{ i}) \,.
\ee
Here the scale range is taken from $k=0.001\,h/{\rm Mpc}$ to $0.2\,h/{\rm Mpc}$ to avoid the non-linear effect \citep{Huang2012JCAP}.

\begin{figure}
        \centering
	\includegraphics[width=0.47\textwidth]{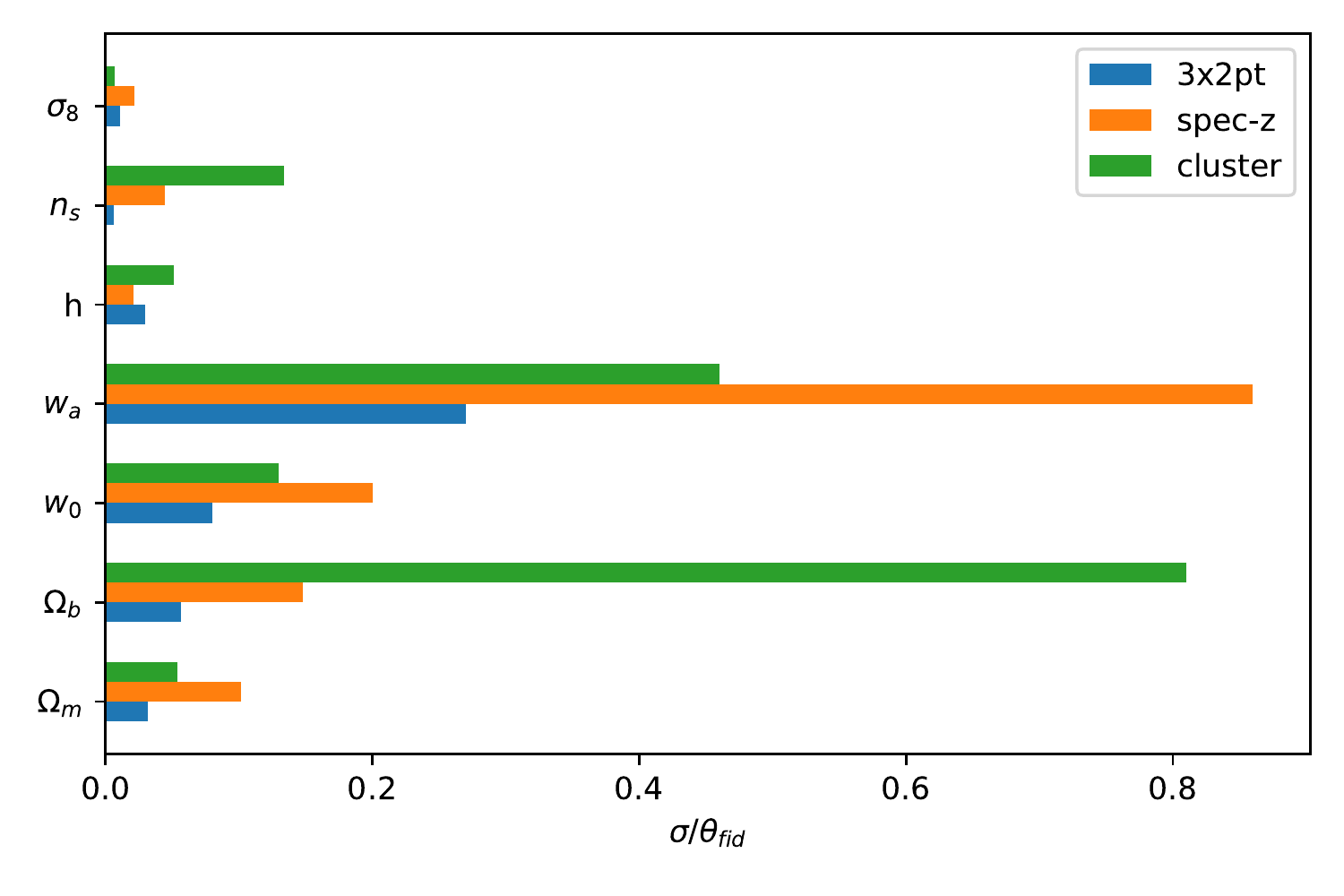}
    \caption{The ratios of 1$\sigma$ errors to the fiducial values of the cosmological parameters for the CSST 3$\times$2pt (blue), spectroscopic galaxy clustering (orange), and cluster number counts (green) surveys, respectively.}
    \label{errors}
\end{figure}

\begin{table*}
\caption{The 1 $\sigma$ errors of cosmological parameters and the ratios of 1$\sigma$ errors to their fiducial values (relative accuracy) for the CSST 3$\times$2pt, Spectroscopic galaxy clustering, cluster number counts, and joint surveys.}
\centering
\begin{tabular}{|c|c| c| c| c| c| c| c|c|c|}
\hline
\hline
Parameter & \multicolumn{1}{c}{fiducial} & \multicolumn{2}{c}{3$\times$2pt analysis} & \multicolumn{2}{c}{spec-z galaxy clustering} & \multicolumn{2}{c}{cluster number counts} & \multicolumn{2}{c}{joint constraint} \\
 & value & 1$\sigma$ error & relative accuracy & 1$\sigma$ error & relative accuracy & 1$\sigma$ error & relative accuracy & 1$\sigma$ error & relative accuracy \\
\hline
$\Omega_{\rm m}$ & 0.314  & 0.01 & 3\% & 0.032 & 10\% & 0.017 & 5\% & 0.004 & 1\% \\
$\Omega_{\rm b}$ & 0.0494 & 0.0028 & 6\% & 0.0073 & 15\% & 0.04 & 80\% & 0.0014 & 2\% \\
$w_0$ & -1 & 0.08 & 8\% & 0.2 & 20\% & 0.13 & 13\% & 0.045 & 4.5\% \\
$w_a$ & 0 & 0.27 & 27\% & 0.86 & 86\% & 0.46 & 46\% & 0.17 & 17\% \\
$h$ & 0.6732 & 0.0199 & 3\% & 0.0143 & 2\% & 0.0344 & 5\% & 0.002 & 0.3\% \\
$n_{\rm s}$ & 0.966 & 0.006 & 0.6\% & 0.0432 & 4\% & 0.129 & 13\% & 0.0039 & 0.4\% \\
$\sigma_{8}$ & 0.82 & 0.009 & 1\% & 0.0177 & 2\% & 0.0059 & 0.7\% & 0.0016 & 0.2\%\\
\hline
\end{tabular}
\label{table3}
\end{table*}

\begin{figure}
        \centering
	\includegraphics[width=0.47\textwidth]{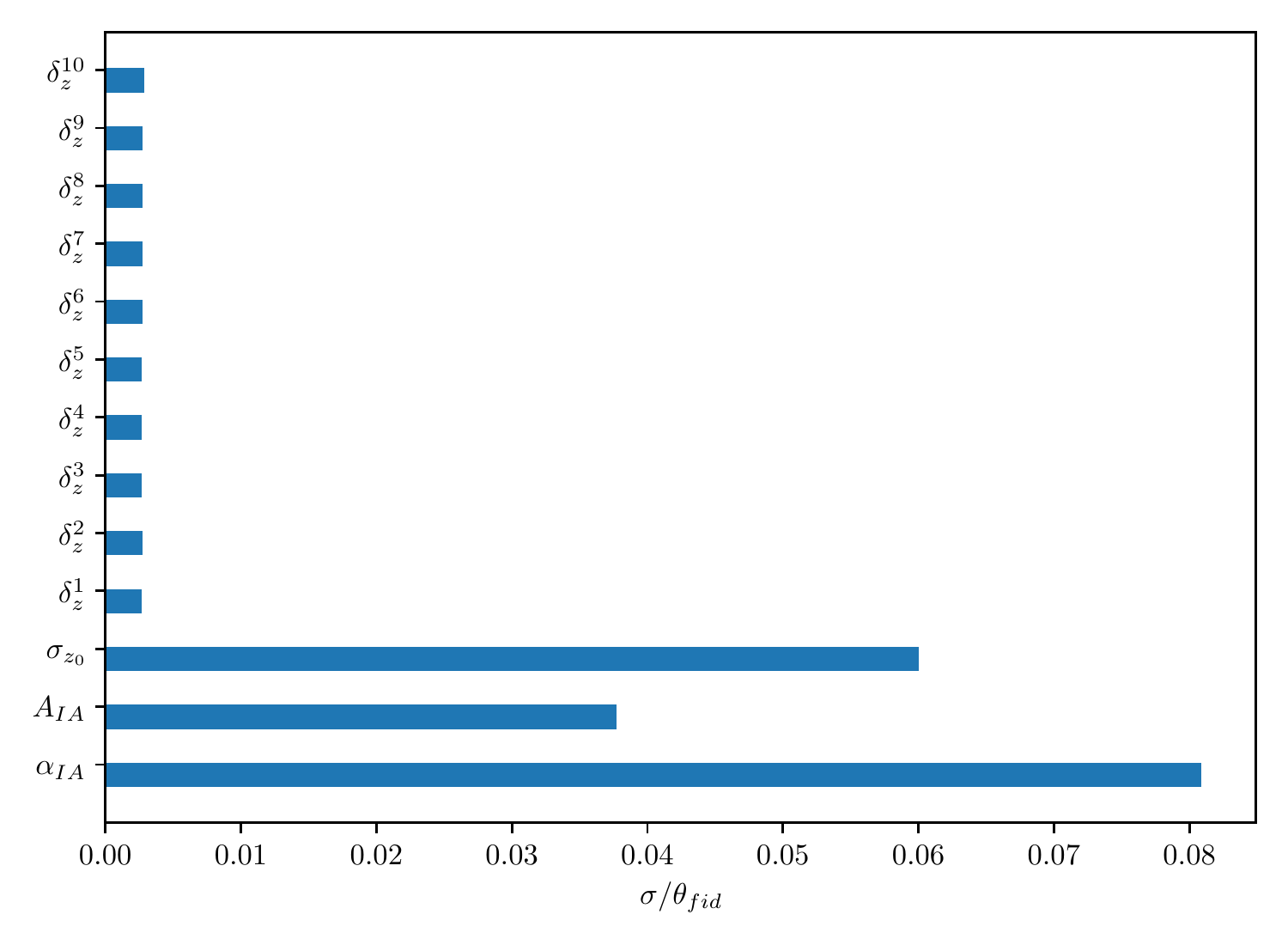}
	\includegraphics[width=0.47\textwidth]{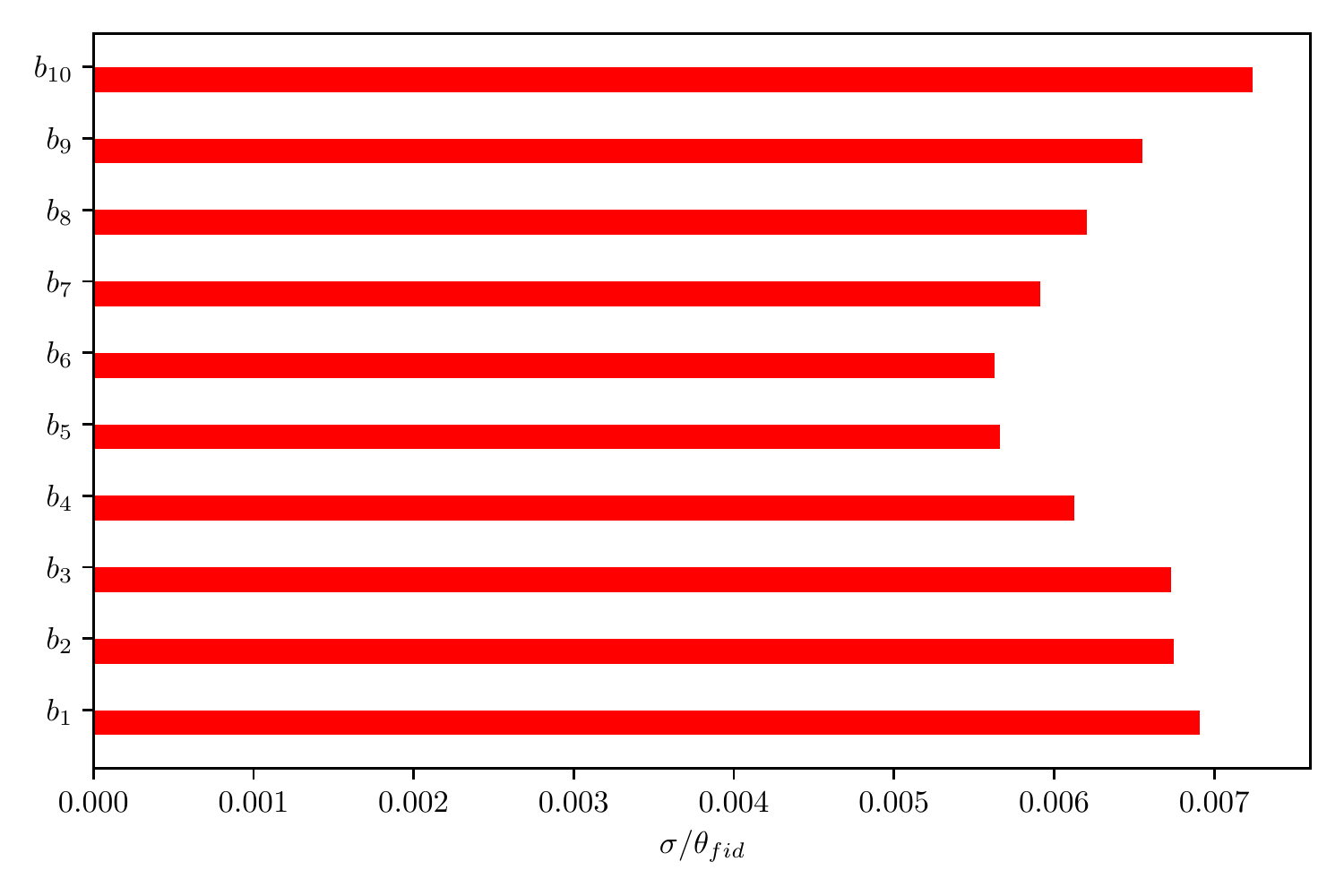}
	\includegraphics[width=0.47\textwidth]{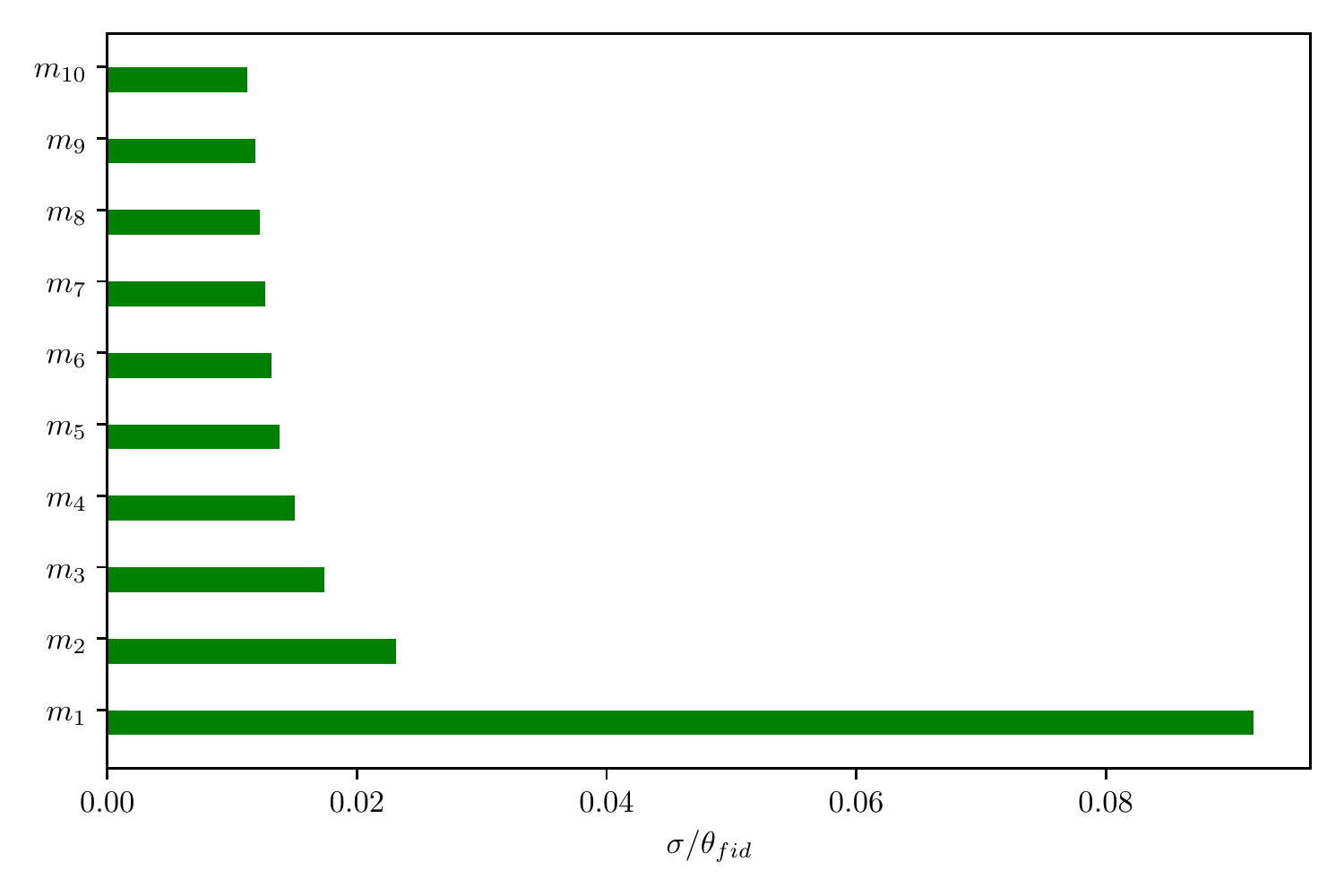}
    \caption{The ratios of 1$\sigma$ errors to the fiducial values of the systematical parameters for the CSST 3$\times$2pt analysis.}
    \label{bias}
\end{figure}

\begin{figure}
        \centering
	\includegraphics[width=0.47\textwidth]{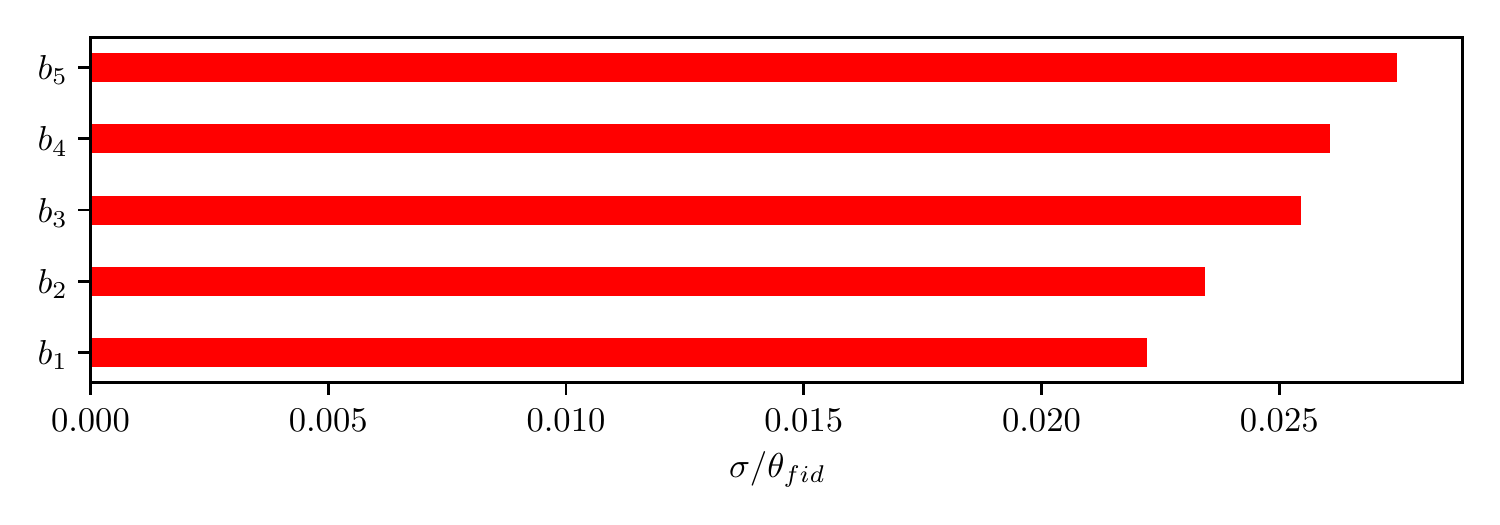}
    \caption{The ratios of 1$\sigma$ errors to the fiducial values of the galaxy biases in the five spec-$z$ bins for the CSST spectroscopic galaxy clustering survey.}
    \label{speczbias}
\end{figure}

\subsection{Forecast for cluster number counts}
The uncertainties of cluster number counts are mainly from the statistical uncertainties of shot noise and sample variance, and the systematical uncertainties in the measurements related to cluster mass determination. For simplicity, we only consider the shot noise term and a constant systematical error that includes all systematics that affect the mass measurement.
In order to calculate Fisher matrix of cluster numbers, we assume the covariance matrix is given by
\be
{\rm Cov} = C^{{\rm SN}}_{\alpha\beta ij} + C_{\rm sys}\,,
\ee
where
\be
C^{{\rm SN}}_{\alpha\beta ij}= N_{\alpha i}\,\delta_{\alpha\beta}\,\delta_{ij}\,,
\ee
and $C_{\rm sys}$ is assumed to be 30\% of $C^{{\rm SN}}_{\alpha\beta ij}$ or the number of galaxy clusters in a redshift bin $\alpha$ and mass bin $i$. This constant systematics are estimated by the current photometric surveys \citep[see e.g.][]{Costanzi2019}, which should be conservative for the CSST photometric and spectroscopic galaxy cluster surveys.
The Fisher matrix then is written as
\be
F_{\theta_{m} \theta_{n}} = \sum_{\alpha i}\frac{\partial N_{\alpha i}}{\partial \theta_{m}}{\rm Cov^{-1}}\frac{\partial N_{\alpha i}}{\partial \theta_{n}}\,.
\ee
\section{Results and Discussion}\label{discussion}

We present the marginalized 1$\sigma$ and $2\sigma$ (i.e. 68.3\% and 95.4\% confidence levels, respectively) contour maps and probability distribution functions (PDFs) of the seven cosmological parameters in Fig.~\ref{all} and \ref{combined}. Fig.~\ref{all} shows the results from CSST 3$\times$2pt, spectroscopic galaxy clustering and cluster number counts surveys, respectively, and Fig.~\ref{combined} shows the joint fitting results with all of these three probes. The joint fitting results are calculated by using the total Fisher matrix, which is given by $F_{\rm{tot}} = F_{\rm{3\times2pt}} + F_{\rm{spec-z}} + F_{\rm{cluster}}$. We have marginalized all the systematical parameters to obtain these contour maps. The details of the forecast results of 1$\sigma$ error and relative accuracy (the ratio of 1$\sigma$ error to the fiducial value) for the seven cosmological parameters with different data sets are listed in Table.\,\ref{table3}. The ratios of 1$\sigma$ errors to the fiducial values of the seven cosmological parameters for the CSST 3$\times$2pt, spectroscopic galaxy clustering, and cluster number counts surveys are shown in Fig.~\ref{errors} for comparison.

In the 3$\times$2pt analysis, we find that the 1$\sigma$ errors (and relative accuracies) of the constraint results on $w_{\rm{0}}$, $w_{\rm{a}}$, $\Omega_{\rm m}$ and $\sigma_8$ are given by $\sigma_{w_{\rm{0}}} = 0.08$ (8\%), $\sigma_{w_{\rm{a}}} = 0.27$ (27\%), $\sigma_{\Omega_{\rm m}}$ = 0.01 (3\%) and $\sigma_{\sigma_8}$ = 0.009 (1\%), respectively. Comparing to the present photometric surveys, e.g. DES \citep{Abbott2022}, our constraints on the dark matter related parameters $\Omega_m$ and $\sigma_8$ are improved by a factor of 3-5, and a factor of $\sim$4 better for the equation of state of dark energy. This is due to that the CSST photometric surveys have larger survey area, deeper magnitude limit, and wider wavelength coverage with seven bands, which could significantly reduce the uncertainties from cosmic variance, Poisson noise, and photo-$z$ calibration. Besides, CSST has higher imaging quality with a high spatial resolution $\sim$0.15'' (80\% energy concentration region) and a Gaussian-like PSF \citep{Gong19,Zhan2021}. All of these advantages can be helpful to significantly improve the constrains on the cosmological parameters. Note that we do not consider the magnification effects here, which may lead to large biases but negligible errors for the constraint results \citep{Unruh2020,Thiele2020,Duncan2021,Lepori2021,Mahony2021}.

For the CSST spectroscopic galaxy clustering survey, we get $\sigma_{w_{\rm{0}}} = 0.2$ (20\%), $\sigma_{w_{\rm{a}}} = 0.86$ (86\%), $\sigma_{\Omega_{\rm m}}$ = 0.032 (10\%) and $\sigma_{\sigma_8}$ = 0.0177 (2\%). Compared to the current spectroscopic surveys, e.g. BOSS \citep{Alam21}, our constraint results of these cosmological parameters are improved by  factors of $\sim$2-5 at least. This is mainly because of larger survey area and deeper magnitude limit in the CSST spectroscopic surveys, that much larger effective survey volume can be obtained to effectively reduce the statistical uncertainties. Although the spectral resolution in the CSST slitless spectroscopic survey is relatively low ($R\simeq200$), which may lead to a low spec-$z$ accuracy, it would not significantly affect the 3D galaxy clustering measurements in the linear regime, given that the spec-$z$ accuracy is expected to be able to achieve 0.002-0.003 using the joint analysis with the CSST photometric data.

In the CSST cluster number counts survey, we obtain $\sigma_{w_{\rm{0}}} = 0.13$ (13\%), $\sigma_{w_{\rm{a}}} = 0.46$ (46\%), $\sigma_{\Omega_{\rm m}}$ = 0.017 (5\%) and $\sigma_{\sigma_8}$ = 0.0059 (0.7\%). Comparing to the results from DES and SPT \citep{Costanzi2021}, our constraints on $\Omega_{\rm m}$ and $\sigma_8$ are improved by 3 and 6 times, respectively. The improvement of constraint on the dark energy equation of state is expected to be in the same order. Here we should note that we assume a perfect mass-richness relation, that the uncertainties from redshift measurement is ignored since galaxy spec-$z$ can be measured in the CSST spectroscopic survey. Besides, other systematics, such as cluster miscentering, cluster projections, and cluster triaxiality \citep{Simet2017,Sunayama2020,Zhang2022,Wu2022}, are also neglected in our analysis. In addition, we know that the abundance of galaxy clusters and the clustering of galaxies are not independent cosmological probes, since galaxy clusters represent the high peaks of galaxy density field. Our current forecast ignores their correlations, and may result in a relatively optimistic estimation. In the future work, we will  study these effects in detail.

In Fig.~\ref{all}, we can see that the degeneracy directions of the cosmological parameters are different for the different three probes. Hence, performing a joint constraint by including the CSST 3$\times$2pt analysis, spectroscopic galaxy clustering, and cluster number counts surveys will provide much more precise constraint results on the cosmological parameters. As shown in Fig.~\ref{combined}, the 1 $\sigma$ errors from the joint constraint for $w_0$, $w_a$, $\Omega_{\rm m}$ and $\sigma_8$ are given by $\sigma_{w_{\rm{0}}} = 0.045$ (4.5\%), $\sigma_{w_{\rm{a}}} = 0.17$ (17\%), $\sigma_{\Omega_{\rm m}}$ = 0.004 (1\%) and $\sigma_{\sigma_8}$ = 0.0016 (0.2\%). We can see that the joint constraint results are significantly improved by factors of 2-5 compared to the results from the three single CSST cosmological probes.


Besides the cosmological parameters, we also show the ratios of 1$\sigma$ errors to fiducial values of the systematical parameters in the three CSST cosmological probes in Fig.~\ref{bias} and Fig.~\ref{speczbias}. In  Fig.~\ref{bias}, we can see that the constraints on photo-$z$ bias and scatter parameters $\delta_z^i$ and $\sigma_{z0}$ are smaller than $\sim$0.3\% and 6\%, respectively, galaxy bias parameters $b_i$ are less than $\sim$0.7\%, and most of the multiplicative shear bias are less than $\sim$2\%. The intrinsic alignment parameters $A_{\rm IA}$ and $\alpha_{\rm IA}$ can be constrained within $\sim$4\% and 8\% accuracies, respectively. We also present the ratios of the 1$\sigma$ errors to the fiducial values for the five galaxy biases in the CSST spectroscopic survey in Fig.~\ref{speczbias}, and we find the constraint accuracy can be less than $\sim$3\%. These constraint results of the galaxy biases are derived from the constraints on the redshift-dependent parameters (see Subsection~\ref{fisher_spec}). The contour maps of the systematical parameters for the 3$\times$2pt analysis and spectroscopic galaxy clustering survey can be found in the Appendix. These results indicate that the CSST photometric and spectroscopic cosmological surveys can simultaneously put strong constraints on both cosmological and systematical parameters.

\section{Summary and Discussion}\label{conclusion}

In the next few years, several space- and ground-based telescopes will be performed and put into service. As one of them with a wide field of view and powerful observational capability, CSST will become an excellent instrument for the cosmological probes. In this paper, we explore the CSST photometric and spectroscopic multi-probe cosmological surveys, including the 3$\times$2pt analysis, spectroscopic galaxy clustering, and galaxy cluster number counts.

We firstly simulate the mock data of the CSST photometric surveys, including weak lensing, photometric 2D galaxy clustering, and galaxy-galaxy lensing surveys, i.e. 3$\times$2pt analysis. The systematical effects from intrinsic alignment, photo-$z$ calibration, galaxy bias, multiplicative error in the shear calibration, and instrument are also considered. The photo-$z$ range is divided into 10 tomographic bins to extract more information. Secondly, we generate mock catalog and galaxy redshift distribution of the CSST spectroscopic galaxy clustering survey based on the zCOSMOS survey. The systematics from galaxy bias and instrument are included in the analysis. The spec-$z$ range is divided into 5 bins to investigate the redshift evolution of related cosmological parameters, such as the equation of state of dark energy. Thirdly, the CSST number counts of galaxy clusters are explored. The mock data and cluster redshift distribution are created based on the halo mass function, and the redshift range is restricted within $z=1.5$ to effectively include both CSST photometric and spectroscopic observations. The cluster redshift distribution is divided in to 10 bins to explore the evolution effects.

The Fisher matrix is employed to analyze and predict the constraints on the cosmological and systematical parameters. We find that all of the three CSST cosmological probes can provide much more stringent constraints on both of cosmological and systematical parameters, that can be improved by several times at least compared to the current corresponding surveys. This is mainly due to the larger survey area, deeper magnitude limit, wider wavelength coverage, and higher imaging quality of the CSST surveys. After combining all data sets to perform the joint constraints, we find that the results can be further significantly improved, which can achieve the constraint accuracies of $\Omega_{\rm m}$ and $\sigma_8$ less than 1\%, and $w_0$ and $w_a$ less than $5\%$ and $20\%$, respectively. The systematical parameters also can be simultaneously constrained within 1\%-10\% accuracy. These results indicate that CSST could be a powerful telescope to explore our Universe, and would greatly improve the studies on important cosmological problems.

Comparing CSST to other next-generation survey telescopes, such as Euclid and LSST, they have similar scientific goals in cosmology and good complementarities and synergies on performances. All of them will launch or operate around 2023-2024, and measure weak gravitational lensing and galaxy clustering to study dark energy and dark matter. Euclid space telescope will observe 15000 deg$^2$, which has large overlapping sky area covered by CSST. Since Euclid only has one single optical broadband covering 550-900 nm, it cannot accurately measure photo-$z$. Then CSST will be very helpful to provide accurate photo-$z$ information with its seven photometric bands from NUV to NIR for Euclid \citep{Cao2018,Zhou2021,Zhou2022}. On the other hand, Euclid has three near-infrared bands from 920-2000 nm, i.e. $Y$, $J$, and $H$, which can significantly improve the accuracy of photo-$z$ and shear measurements in CSST surveys \citep{Cao2018,Liu2022}. As a ground-based telescope, LSST will explore 18000 deg$^2$ with deeper magnitude limit compared to CSST and Euclid, that can obtain more faint and high-$z$ galaxy samples for cosmological studies. Besides, a few hundred thousand Type Ia supernovae (SNe Ia) probably can be detected by LSST per year. This would be a great complement to the study of cosmic expansion history in CSST and Euclid surveys, that only a few thousand $\rm SNe\ Ia$ are expected to be measured in their mission durations. On the other hand, CSST and Euclid have much better spatial resolution, and can measure smaller galaxies and obtain better galaxy shape measurement in weak gravitational lensing survey compared to LSST. Hence, there are large complementarities between CSST and other future survey telescopes. Synergies of these experiments can provide extremely accurate constraint on cosmological parameters, and are powerful to reveal the nature of dark energy and dark matter.

\section*{Acknowledgements}

H.T.M. and Y.G. acknowledge the supports from 2020SKA0110402, MOST-2018YFE0120800, NSFC-11822305, NSFC-11773031, and NSFC-11633004. X.L.C. acknowledges the support of the National Natural Science Foundation of China through grant Nos. 11473044 and 11973047, and the Chinese Academy of Science grants QYZDJ-SSW-SLH017, XDB 23040100, and XDA15020200. Z.Q.H acknowledges the support of the National key R\&D Program of China (Grant No. 2020YFC2201600), National SKA Program of China No. 2020SKA0110402, National Natural Science Foundation of China (NSFC) under Grant No. 12073088, and Guangdong Major Project of Basic and Applied Basic Research (Grant No. 2019B030302001). X.D.L acknowledges the support from the NSFC grant (No. 11803094) and the Science and Technology Program of Guangzhou, China (No. 202002030360). This work is also supported by science research grants from the China Manned Space Project with grants Nos. CMS-CSST-2021-B01 and CMS-CSST-2021-A01.

\section*{Data Availability}

 The data that support the findings of this study are available from the corresponding author, upon reasonable request.


\bibliographystyle{mnras}
\bibliography{multra} 


\appendix

\section{contour maps of the systematical parameters}

\begin{figure*}
        \centering
	\includegraphics[width=\textwidth]{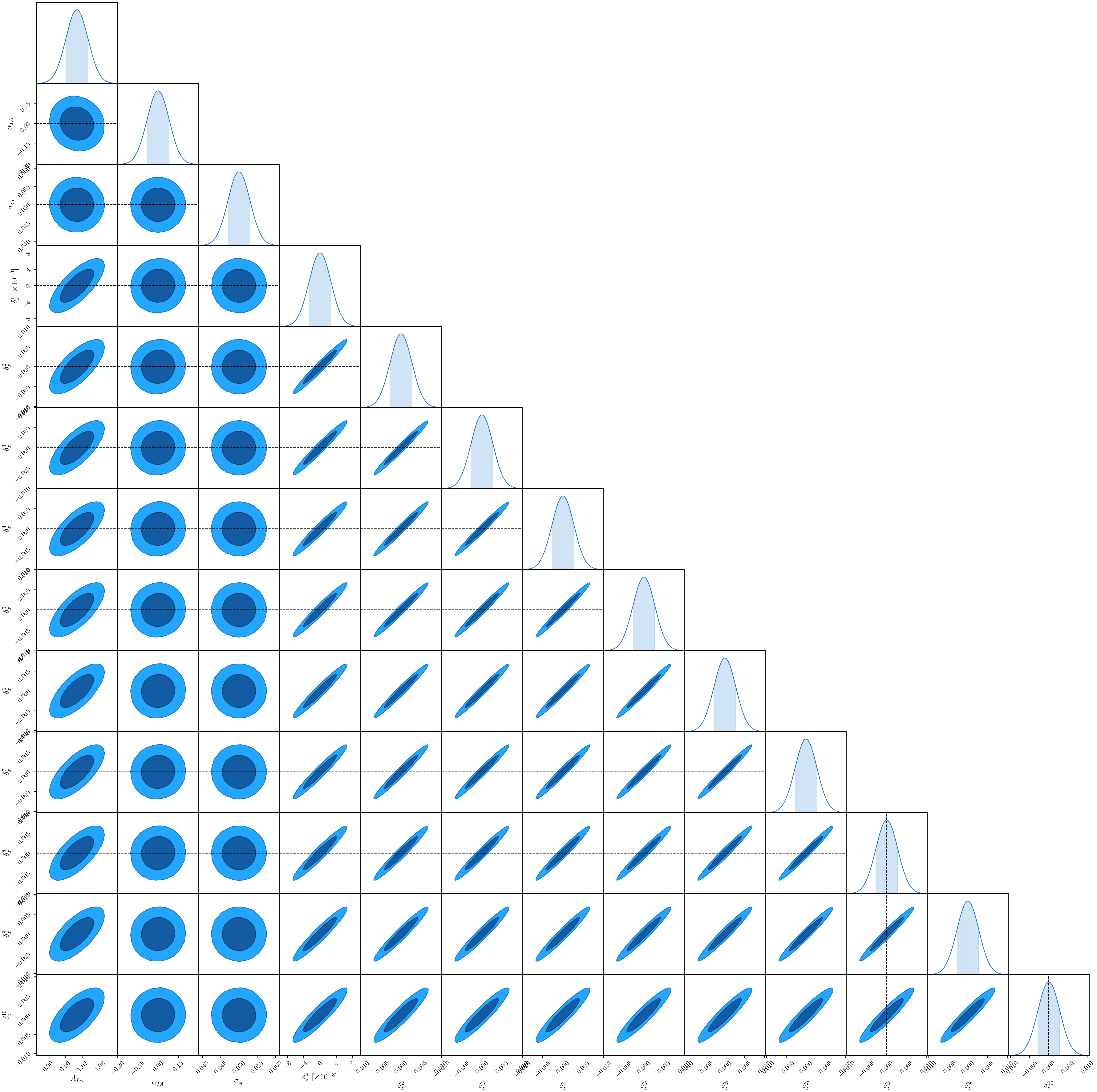}
    \caption{Constraint results of the parameters for the IA and photo-$z$ calibration in the CSST 3$\times$2pt probes.}
    \label{photobias}
\end{figure*}

\begin{figure*}
        \centering
	\includegraphics[width=1.0\textwidth]{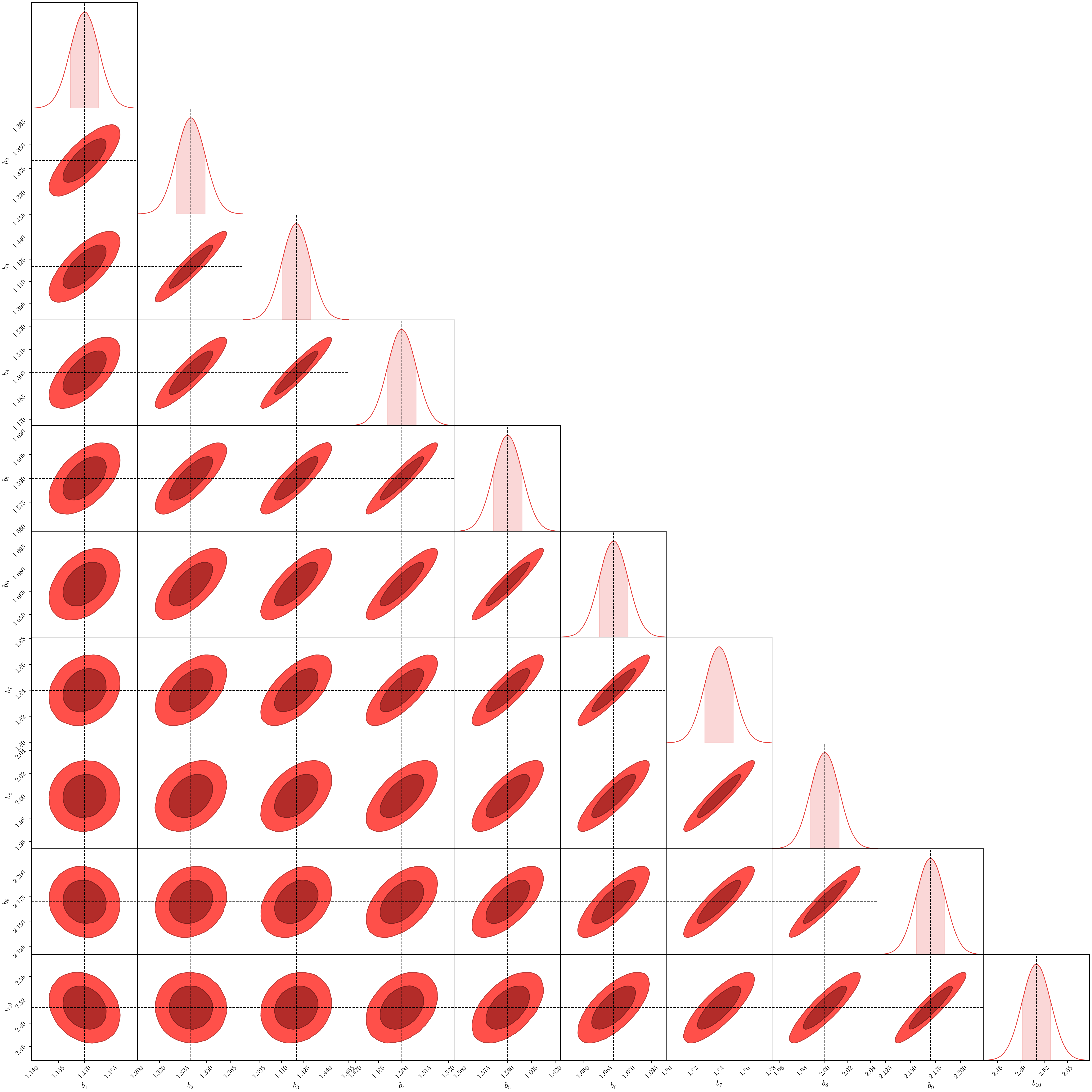}
    \caption{Constraint results of the galaxy biases in the 10 photo-$z$ bins of the CSST 3$\times$2pt probes.}
    \label{galaxybias}
\end{figure*}

\begin{figure*}
        \centering
	\includegraphics[width=1.0\textwidth]{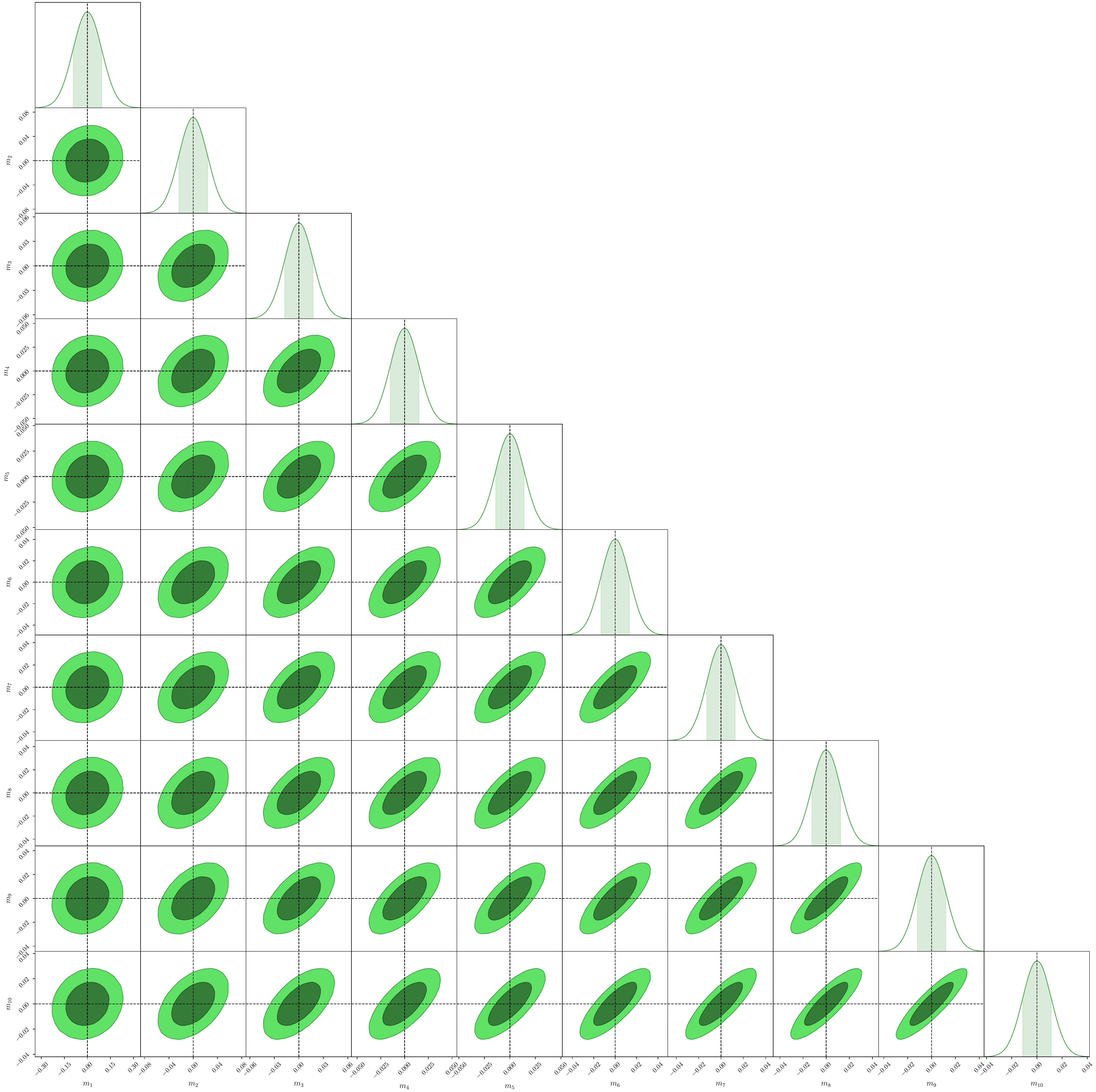}
    \caption{Constraint results of the shear multiplicative biases in the 10 photo-$z$ bins of the CSST 3$\times$2pt probes.}
    \label{shearbias}
\end{figure*}

\begin{figure*}
        \centering
	\includegraphics[width=1.0\textwidth]{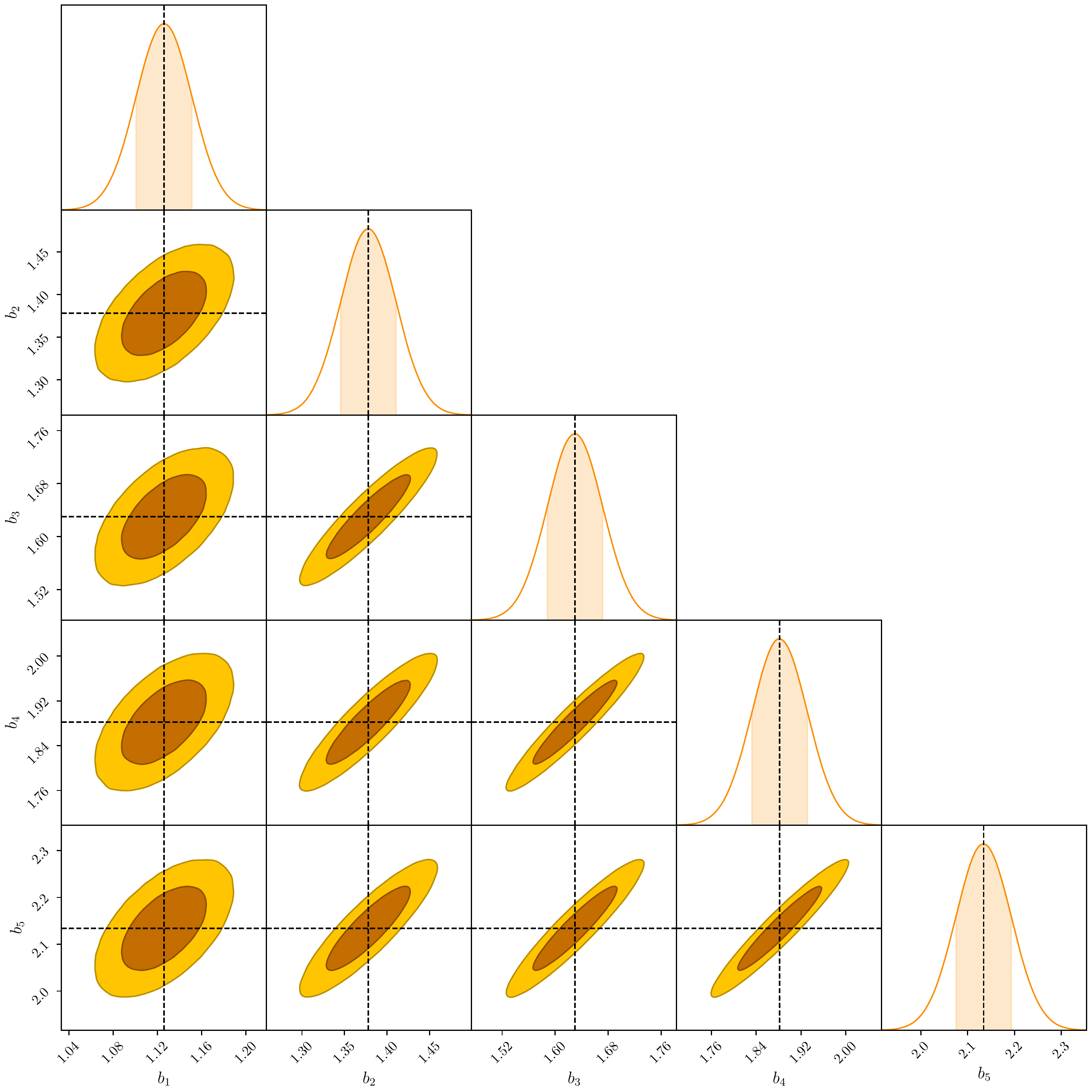}
    \caption{Constraint results of the galaxy biases in the 5 spec-$z$ bins of the CSST spectroscopic galaxy clustering survey.}
    \label{specbias}
\end{figure*}

The contour maps of the systematical parameters in the CSST 3$\times$2pt probes, i.e. IA and photo-$z$ calibration, galaxy bias, and multiplicative error, are shown in Fig.~\ref{photobias}, \ref{galaxybias}, and \ref{shearbias}, respectively. The constraint results of the galaxy biases in the CSST spectroscopic galaxy clustering survey are also presented in Fig.~\ref{specbias}.

\bsp	
\label{lastpage}
\end{document}